\documentclass[]{PoS}
\usepackage{cite}

\title{Evidence for a hadronic origin of the Fermi Bubbles and the  Galactic Excess}

\ShortTitle{Evidence for a hadronic origin of the Fermi Bubbles and the Central GeV excess}

\author{\speaker{Wim de Boer}\\
        Institut f\"ur Experimentelle Kernphysik, Karlsruhe Institute of Technology, P.O. Box 6980, 76049 Karlsruhe, Germany\\
        E-mail: \email{wim.de.boer@kit.edu}}

\author{Iris Gebauer, Simon Kunz, Alexander Neumann\\
        Institut f\"ur Experimentelle Kernphysik, Karlsruhe Institute of Technology, P.O. Box 6980, 76049 Karlsruhe, Germany\\
        E-mail: \email{Iris.Gebauer@kit.edu;simon.kunz@kit.edu;alexander.neumann2@student.kit.edu}}

\abstract{The Fermi-LAT gamma-ray  data revealed giant 'bubbles' of emission above and below the Galactic plane   with an energy spectrum significantly harder than seen from other  directions.  How the bubbles connect to the Galactic plane is unclear. Previous analyses masked the Galactic plane because of the large foreground.  In this paper we use a novel spectral template fit, which allows a simultaneous determination of the foreground and the hard bubble-like emission in any direction.  We find that  bubble-like emission  is not only found in the halo, but it is strongly present in the Galactic plane as well with a morphology close to the spatial distribution of the 1.8 MeV gamma-ray line from $^{26}$Al, a radioactive nucleus synthesized in SNRs.  In addition,
the spectral shape of this hard component  coincides with the predicted spectrum from  cosmic rays trapped in sources (SCRs)  Hence, we  propose that the bubble-like emission in the plane has a hadronic origin, which arises from SCRs. The bubbles in the halo have the same energy spectrum, which suggests that they are outflows from the plane with the gamma-rays arising from hadronic interactions of protons trapped  in a plasma of advected gas.   Evidence for advected gas  is provided by the ROSAT X-ray observations from hot gas in the bubble region. Alternatively, the protons may be accelerated in the shock wave of the outflow, thus having the same spectrum as the SCRs, which are accelerated in shock waves as well.

Towards the Galactic center (GC) we observe the 1-3 GeV excess, but observe this excess in addition in all regions where there is strong $^{26}$Al production.  This excludes the dark matter annihilation interpretation. Instead, we propose that the excess is caused  by a deficit of the CRs at low rigidities in the regions of dense molecular clouds, which are characterized by higher energy losses and stronger stellar winds.   The $^{26}$Al  line is a tracer of such regions.   If we introduce a break in  the cosmic ray injection spectra  in the regions of strong $^{26}$Al production to effectively take the deficit of low rigidity CRs into account,  the excess  in the Galactic plane and towards the GC both disappear. We  thus conclude  that the  excess in the Galactic center is largely  an artefact from the excess in the Galactic plane. The correlation between the two originates simply from the lines-of-sight towards the GC crossing the Galactic plane. }

\FullConference{The 34th International Cosmic Ray Conference,\\
		30 July- 6 August, 2015\\
		The Hague, The Netherlands}

\begin{document}

\section{Introduction}\label{intro}
The FERMI-LAT gamma-ray telescope \cite{Atwood:2009ez} has surveyed the  gamma-ray sky  at energies between 100 MeV and 100   GeV or even above with unprecedented precision.
The main contributions  to the gamma-rays are well understood: cosmic rays (CRs)  interacting with the gas and the photons in the Galaxy, which leads to $\pi^0$ production by nuclear interactions combined with a smaller leptonic component: bremsstrahlung (BR) and inverse Compton  scattering (IC) of electrons on the photons of the interstellar radiation field (ISRF) \cite{Strong:2007nh}.   The Fermi  Bubbles were discovered as an excess over the expected background \cite{Su:2010qj}.  The  origin of the bubbles  is unclear and many proposals have been made, ranging from hadronic cosmic rays (CRs) interacting with hot gas in the halo \cite{Crocker:2010dg,Crocker:2014fla}, to star bursts \cite{Biermann:2009td}, to AGN activity in our Galaxy \cite{Guo:2011eg,Guo:2011ip,Yang:2012fy,Cheng:2011xd},  to  dark matter annihilation   \cite{Dobler:2011mk}.

Currently, the morphology of the bubbles has not been examined in the Galactic plane (GP), because of the large foreground, although such a determination could be helpful for the interpretation. A spectral template fit to the Fermi data allows a simultaneous determination of the foreground and  bubble signature in all sky directions, including the GP. The idea is simple:  the energy spectra of the main contributions to the gamma-ray sky are known from accelerator experiments. If combined with the  energy spectra of the CRs and the  ISRF  the gamma-ray spectra can be calculated for each contribution using one of the public propagation codes, like Dragon \cite{Evoli:2008dv} or Galprop \cite{Moskalenko:1998id,Vladimirov:2010aq}. These are convenient tools, since they include the cross sections from accelerator experiments and take the changes in the CR spectra from propagation and energy losses into account.  If the known contributions of gamma-ray production ($\pi^0$ decay, IC and BR) describe the data one should be able to fit the gamma-ray spectra in each sky direction by a linear combination of them.   Additional contributions will be apparent from a poor fit. For every spatial direction, defined as a cone in longitude and latitude (sometimes called pixel),  the data consist of 21 energy bins with only a few free parameters, namely the normalization of each template. This leads to  a strongly constrained fit in each direction. First results have been published elsewhere \cite{deBoer:2014bra}.

The paper has been organized as follows.  Sect. \ref{anal}  describes the data and the template fit, which allows for a fine spatial resolution, even to the level of molecular clouds (MCs), thought to harbor the sources of CRs. Inside these sources\footnote{Mainly  supernova remnants (SNRs), but  other sources, like pulsars, may contribute as well. Since SNRs are expected to be dominant, we refer to sources as SNRs.} the density of CRs and the gas density in the shock wave are high, so the gamma-ray production from the CRs inside the sources,  called "Source CRs" (SCRs) in Ref. \cite{2004ApJ...611...12B}, may be significant. Since these correspond to gamma-rays from unpropagated CRs they should reveal a hard spectrum expected for shock wave acceleration by SNRs  with a spectral index between 2.0 and 2.3.\cite{Hillas:2005cs}  Evidence for these SCRs is presented and its connection with the Fermi Bubbles, which, interestingly, have the same spectral shape as the SCRs, will be discussed.  Since this emission happens both in the bubbles and the GP we will call it generically $1/E^{2.1}$ emission.

In Sect. \ref{morph}  the morphology of the $1/E^{2.1}$ emission is determined.  In Sect. \ref{ener} the energy of the this emission is compared with the gamma-ray energy generated by the averaged rate of SNRs in our Galaxy, which shows that the rate of SNRs is high enough to generate all $1/E^{2.1}$ emission.
\begin{figure}
\centering
\includegraphics[width=0.42\textwidth,height=0.33\textwidth]{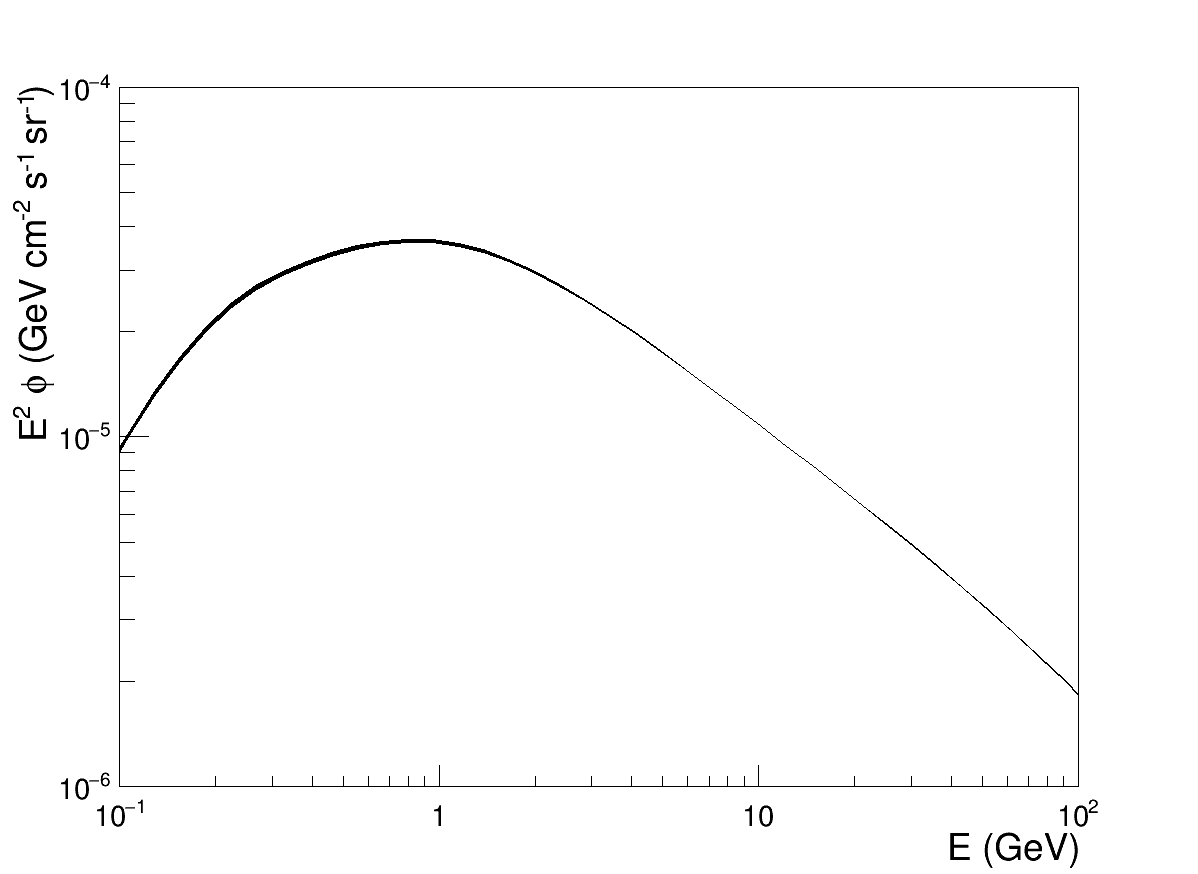}
\hspace*{4mm}
\includegraphics[width=0.42\textwidth,height=0.33\textwidth]{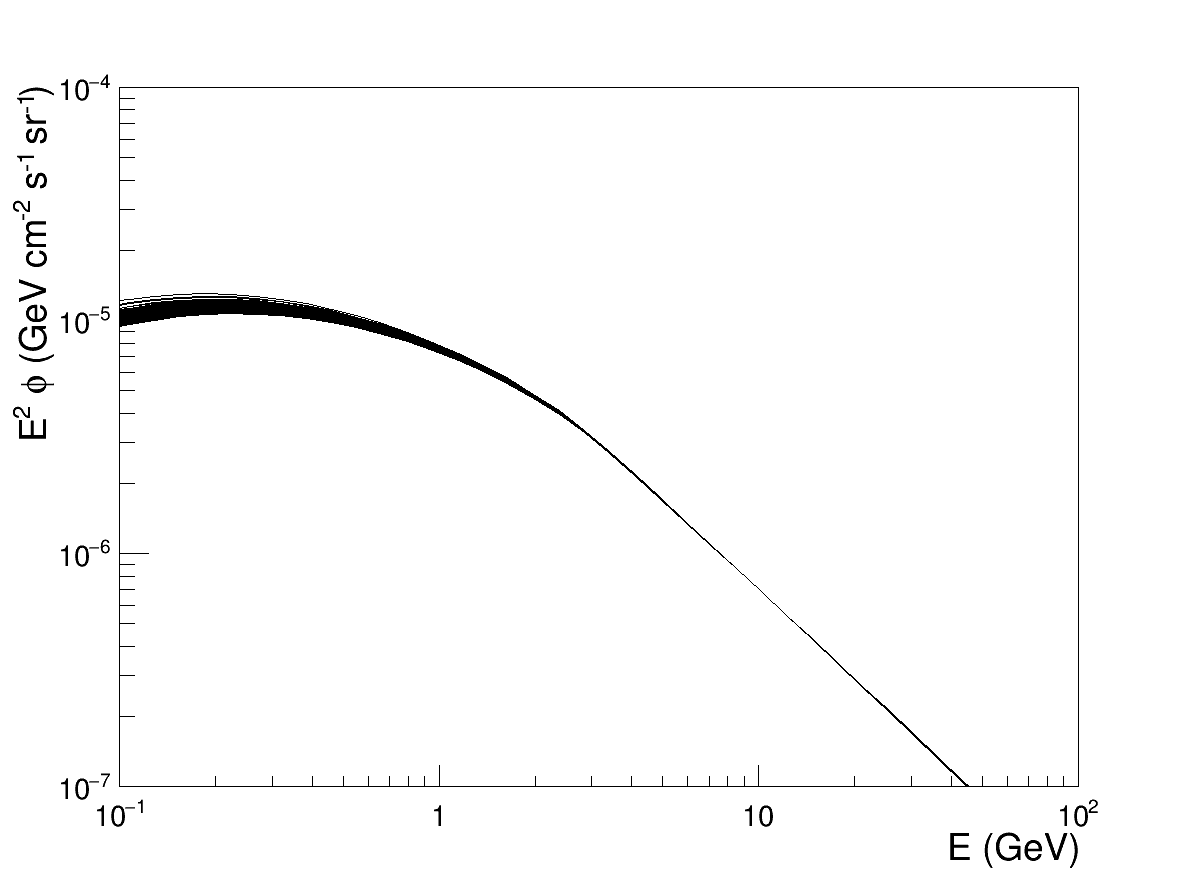}
\hspace*{0.05\textwidth}(a)\hspace*{0.5\textwidth} (b)\\
\includegraphics[width=0.42\textwidth,height=0.33\textwidth]{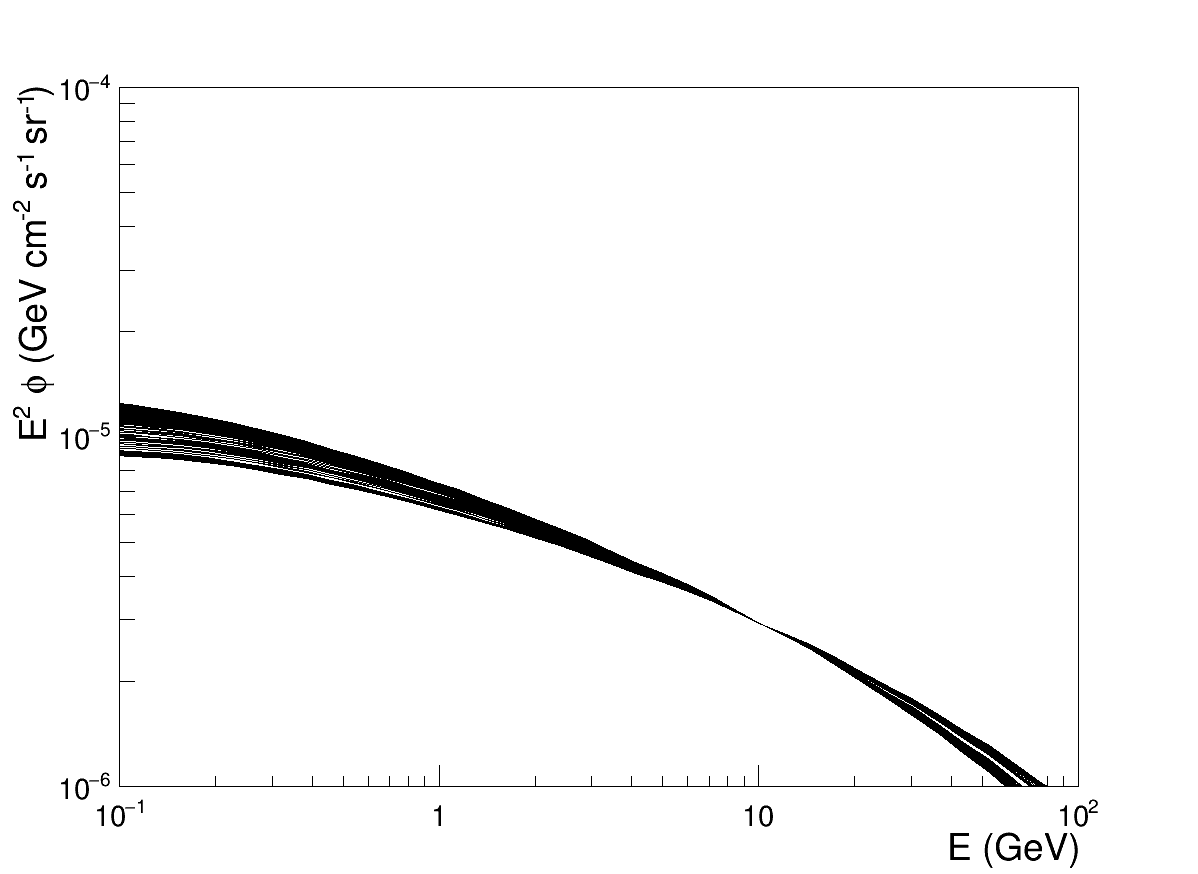}
\hspace*{4mm}
\includegraphics[width=0.42\textwidth,height=0.33\textwidth]{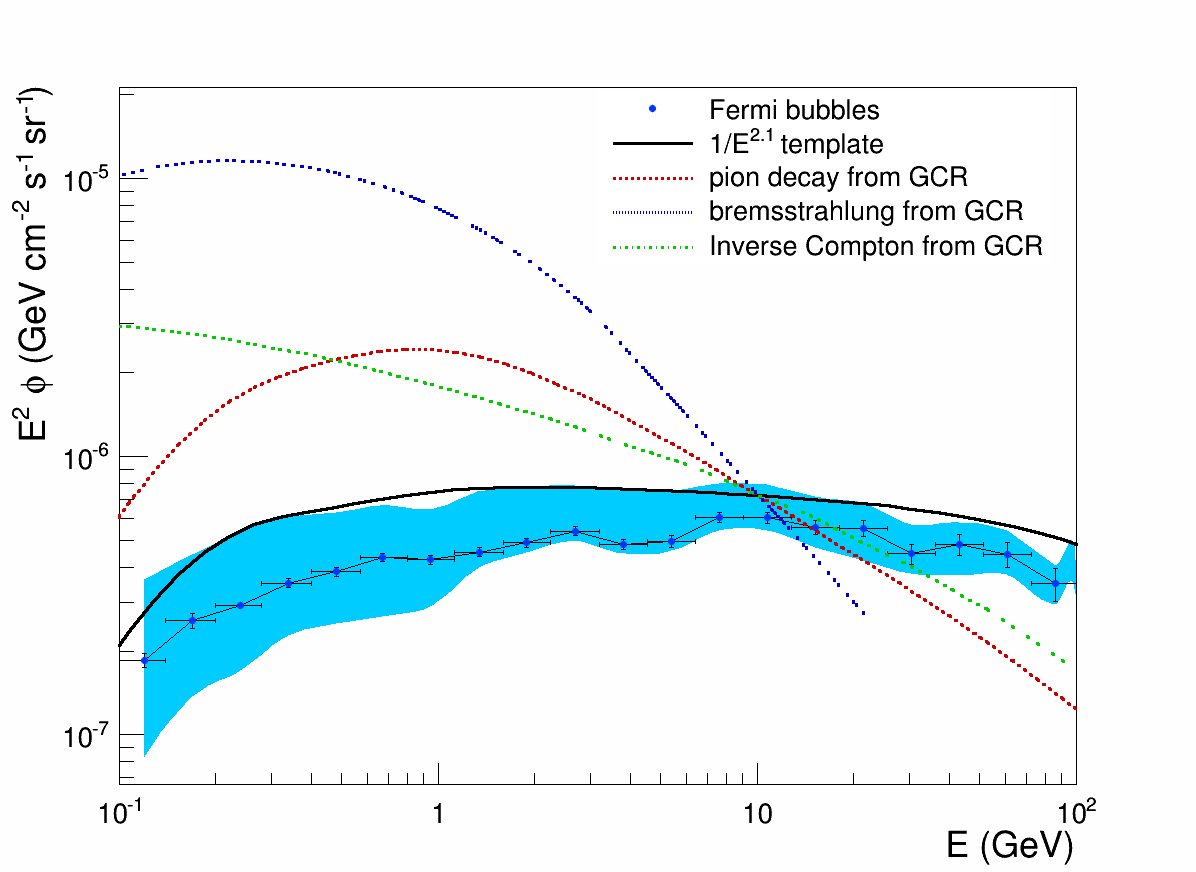}
\hspace*{0.05\textwidth}(c)\hspace*{0.5\textwidth} (d)\\[1mm]
\caption[]{Spectral templates for decays of (a): $\pi^0$ mesons; (b)  Bremsstrahlung; (c):  inverse Compton scattering. Each panel has normalized  templates  superimposed for all considered subcones in the region |b|<60$^\circ$, $|l|<60^\circ$; (d): the Fermi Bubble energy spectrum from Ref. \cite{Fermi-LAT:2014sfa}, as indicated by the data points inside the shaded area. The latter  indicates the spread in the different analyses. The shape of these data is well reproduced by a template  (black line), calculated from   $\pi^0$ production by  protons with a $1/E^{2.1}$ spectrum, as expected for SCRs.  For comparison,  the   $\pi^0$, IC and BR contribution from Galactic CRs (instead of SCRs) are shown as well.  They do not describe the shape of the
  data points. \label{f1}}
\end{figure}
 
In Sect. \ref{26al} we show that the $1/E^{2.1}$  emission has a morphology in the GP identical to  the skymap of the 1.809 MeV line of $^{26}$Al.\cite{Plueschke:2001dc,Kretschmer:2013naa} This radioactive element, which is synthesized in SNRs, is a good tracer of SNRs, since these are expected to be the dominant sources.\cite{Prantzos1996}
The excellent agreement between the two skymaps in the GP, both towards the GC and along the spiral arms, provides convincing evidence that the $1/E^{2.1}$ emission in the GP indeed originates from the  SCRs. The bubbles in the halo are then interpreted as outflows from the GC into the halo with the CRs trapped inside the plasma. These CRs  keep the same spectrum as in the sources, either because the mean free path in the plasma is so short, that the CRs do not escape by energy dependent diffusion or they are accelerated by the shock of the outflow in the halo, thus obtaining the spectrum of diffusively accelerated CRs again.

In Sect. \ref{gce} we confirm the excess in the GC with our template fit. This excess has been widely discussed before. \cite{Goodenough:2009gk,Hooper:2010mq,Boyarsky:2010dr,Morselli:2010ty,Vitale:2011zz,Wharton:2011dv,Hooper:2012sr,YusefZadeh:2012nh,Abazajian:2012pn,Hooper:2013rwa,Mirabal:2013rba,Huang:2013pda,Huang:2013apa,Macias:2013vya,Daylan:2014rsa,Macias:2014sta,Abazajian:2014hsa,Calore:2014xka,Calore:2014nla,Bartels:2015aea,Cholis:2014lta,Cholis:2015dea}
The most exciting interpretations have been the contributions from dark matter (DM) annihilation  and/or unresolved sources, like millisecond pulsars (MSPs). However, with our template fit we observe this excess in every region of the GP with a  strong $^{26}$Al production. Hence, we call it Galacric plane excess (GPE) instead of GCE. If there is a GPE this will show up in the GC as well, since the lines-of-sight towards the GC cross the GP.  A possible interpretation will be discussed. Sect. \ref{conc} summarizes the results.

\section{Analysis}\label{anal}
We have analysed the diffuse gamma-rays in the energy range between 0.1 and 100 GeV using the diffuse class of the public  P7REP\_SOURCE\_V15 data collected from August, 2008 till July 2014 (72 months) by the Fermi Space Telescope \cite{Atwood:2009ez}. The data were analysed  with the recommended selections for the diffuse class using the  Fermi Science Tools (FST)  software \cite{FST}.
The sky maps were binned in longitude and latitude in 0.5x0.5$^\circ$ bins, which could later be combined at will. The point sources  from the second Fermi point source catalogue \cite{Fermi-LAT:2011iqa} have been subtracted using the {\it gtsrc} routine in the FST. The recommended  selection of events allows one to take the events from misidentified hadrons into account, which are part of the isotropic component provided by the Fermi software. 
 
The gamma-ray flux is proportional to the product of the CR densities, the "target densities" (gas or ISRF) and the cross sections, but  a template fit  lumps the product of these three factors into a single normalization factor  for each gamma-ray component $k$, thus eliminating the need to know them individually.

 The total flux in a given direction can be described as a linear combination of the various contributions with known energy spectra (templates):
 \begin{equation} |\Phi_{tot}>=n_1|\Phi_{\pi^0}>\  +\  n_2|\Phi_{BR}>\   +\  n_3|\Phi_{IC}>\  +\  n_4|\Phi_{Bubble}>\  +\  n_5|\Phi_{isotropic}>, \label{e2}\end{equation} where the normalization factors $n_i$ determine the fraction of the total flux for a given contribution. The factors $n_i$, and hence the flux of each contribution, can be found from a $\chi^2$ fit, which tries to adjust the templates to best describe the data.  Since the number of data points in each spectrum for a given subcone (21 energy bins)  is large compared with  the number of free parameters ($n_i\le 5$) the fit is strongly constrained, thus allowing a  determination of the various template contributions in each direction.

As test statistic we use the  $\chi^2$ function defined as:
 \begin{equation} \chi^2=\sum_{i=1}^{N} \sum_{j=1}^{21}\left[\frac{\langle data(i,j)- \sum_{k=1}^{m\le5} n(i,k) \times template(i,j,k)\rangle^2}{\sigma(i,j)^2}\right], \label{e1}\end{equation}
where the sum is taken over the N subcones   in different sky directions $i$, data(i,j) represents the total Fermi flux in direction $i$ for energy bin $j$, template(i,j,k) with normalization $n(i,k)$ is the contribution of template $k$ out of a total of $m$ templates to data(i,j) and  $\sigma(i,j)$ is the total error of data(i,j), obtained by adding the statistical and systematic errors  in quadrature. The recommended systematic errors  in the Fermi Software on the total flux  are 10\% for gamma-ray energies below 100 MeV, 5\% at 562 MeV, and 20\% above 10 GeV. We used  a linear interpolation for energies in between.

\begin{figure}
\centering\vspace*{-0mm}
\includegraphics[width=0.42\textwidth,height=0.33\textwidth,clip]{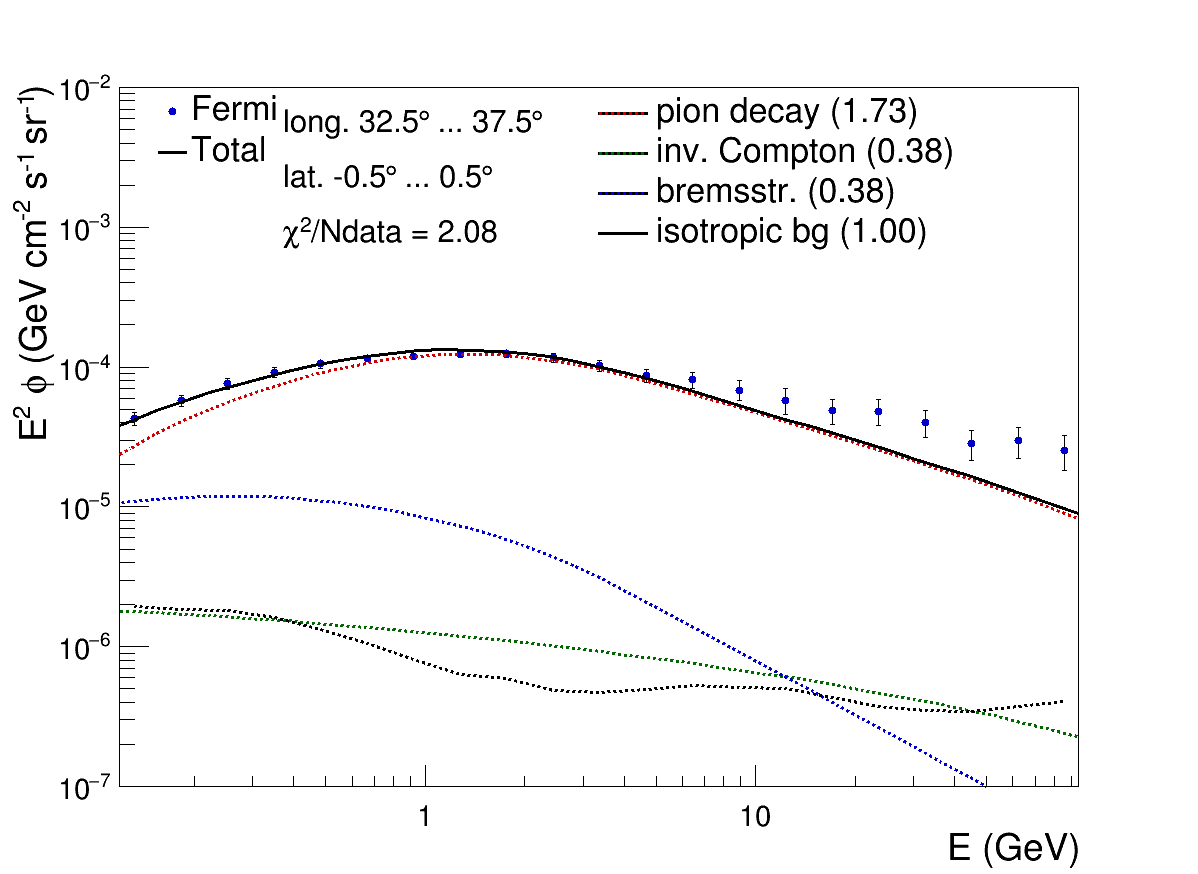}\hspace*{8mm}
\includegraphics[width=0.42\textwidth,height=0.33\textwidth,clip]{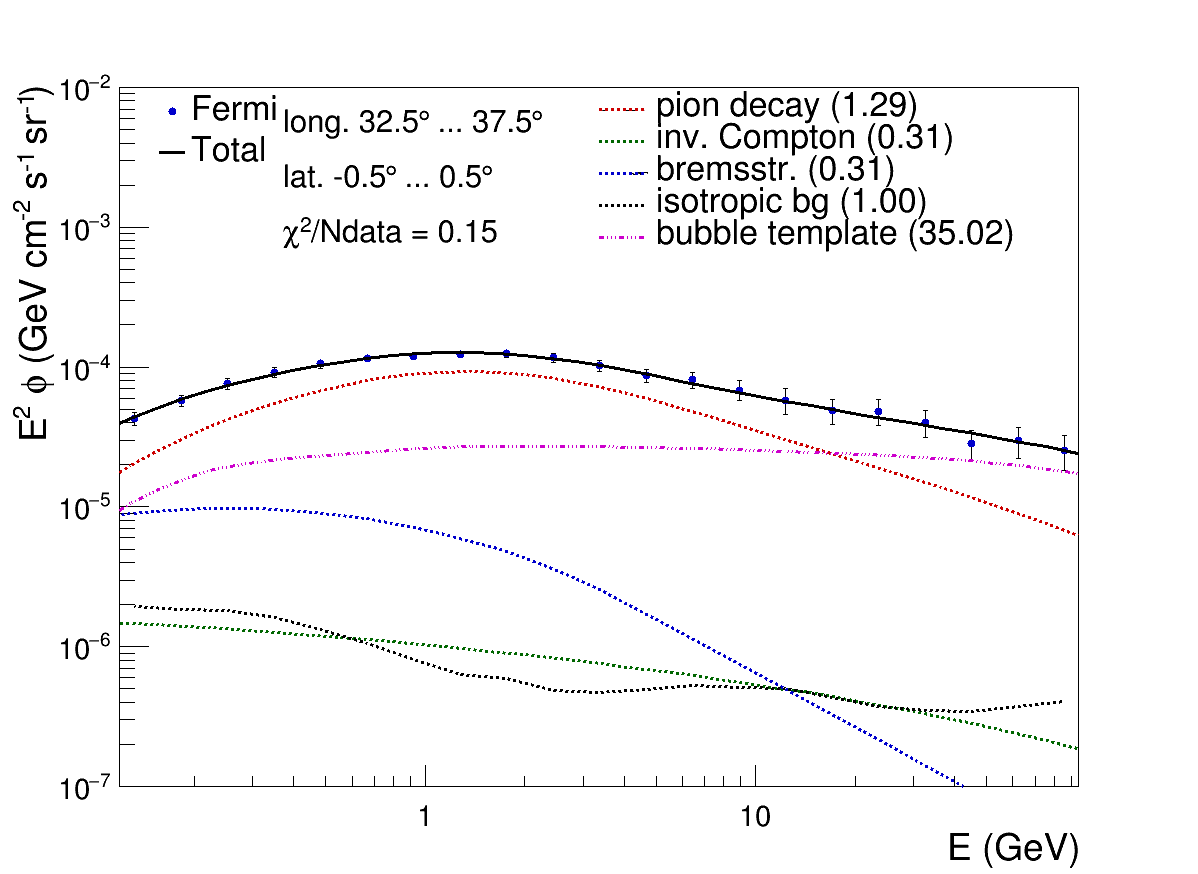}
\hspace*{0.05\textwidth}(a)\hspace*{0.45\textwidth} (b)\\
\hspace*{0mm}\includegraphics[width=0.42\textwidth,height=0.33\textwidth,clip]{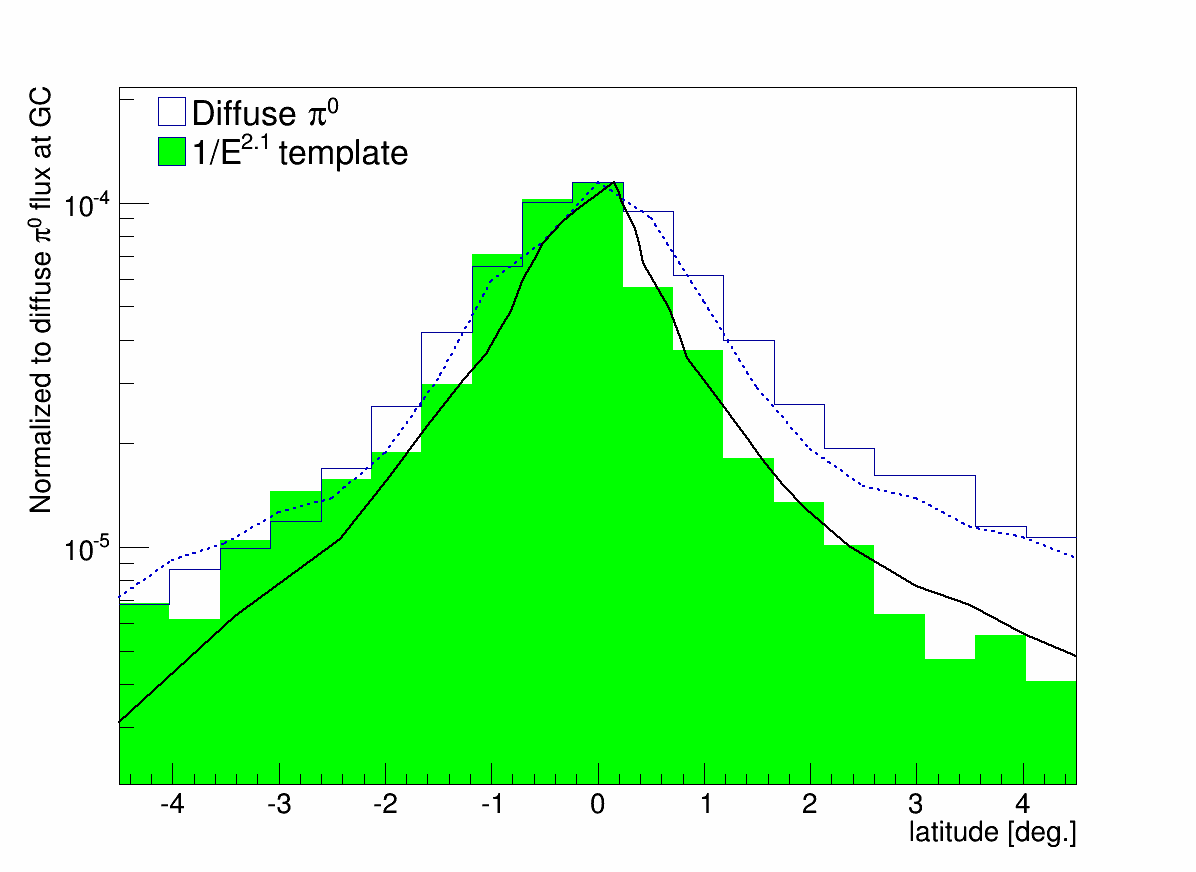}\hspace*{8mm}
\includegraphics[width=0.42\textwidth,height=0.33\textwidth,clip]{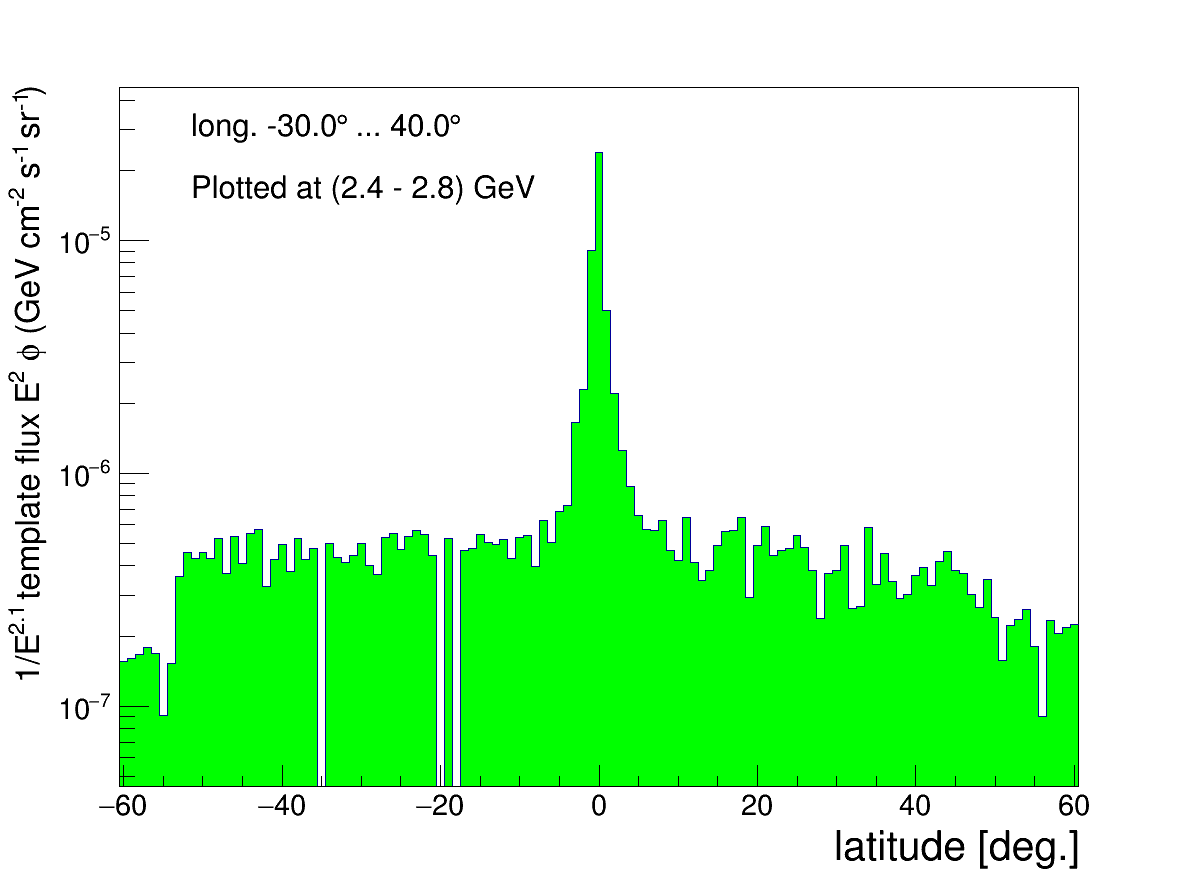}
\hspace*{0.05\textwidth}(c)\hspace*{0.45\textwidth} (d)\\[-0mm]
\caption[]{(a) The results from a template fit without the 1/E$^{2.1}$ template to the gamma-ray spectrum in the GP for longitudes centered around 35$^\circ$. The numbers in brackets in the legend indicate the normalization with respect to the Dragon prediction.
(b) The high energy tail is well described after adding the 1/E$^{2.1}$ template  to the fit.  The strength of the 1/E$^{2.1}$ template is indicated in brackets in units of the black solid curve in Fig. \ref{f1}d. (c) Latitude distributions for the  diffuse $\pi^0$ component  (outer  histogram) and the 1/E$^{2.1}$  emission (inner  green histogram, normalized to the outer histogram at the center)   for $-30^\circ<l<40^\circ$. 
 The dashed line corresponds to the gas distribution as implemented in the Dragon propagation code. The solid line was taken  from the  molecular column density in Ref.  \cite{Dame:2000sp}.
(d)  Latitude distribution of the 1/E$^{2.1}$ template flux at an energy around 2.4 GeV  on a logarithmic scale for $-30^\circ<l<40^\circ$.
 \label{f2}
}
\end{figure}

The foreground templates  depend mainly on the CR spectral shapes. We tuned the CR spectra of electrons and protons and other nuclei to best describe  the gamma-ray sky using the public propagation code Dragon \cite{Evoli:2008dv}.
The resulting foreground templates are shown in Fig. \ref{f1}a-\ref{f1}c. Here we superimposed the templates in  different directions in a cone of 60$^\circ$ around the GC. All spectra were normalized at 3.4 GeV, so only differences in shape, not in flux, are shown. One observes that the $\pi^0$ template does not change with direction, as expected, since the energy losses are small for nuclei.
The Bremsstrahlung  differences are caused by the varying magnetic field in the Galaxy. For IC  the differences originate form the difference in the spectra of the ISRF, which is dominated by the Cosmic Microwave Background (CMB) in the halo, but has contributions from infrared  and stellar light in the GP.  The template differences in different directions were taken into account in the template fit.
\begin{figure}
\centering
\includegraphics[width=0.45\textwidth,height=0.33\textwidth,clip]{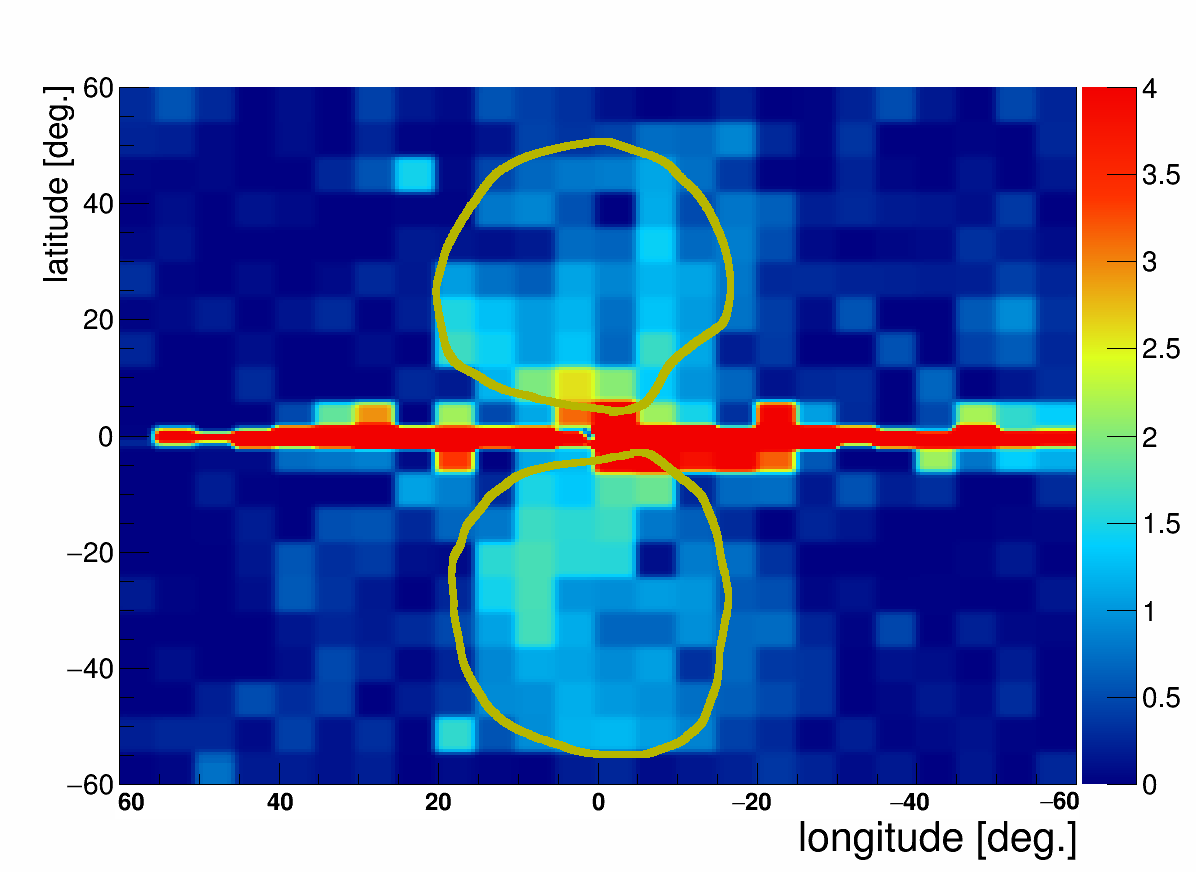}\hspace*{10mm}
\includegraphics[width=0.45\textwidth,height=0.33\textwidth,clip]{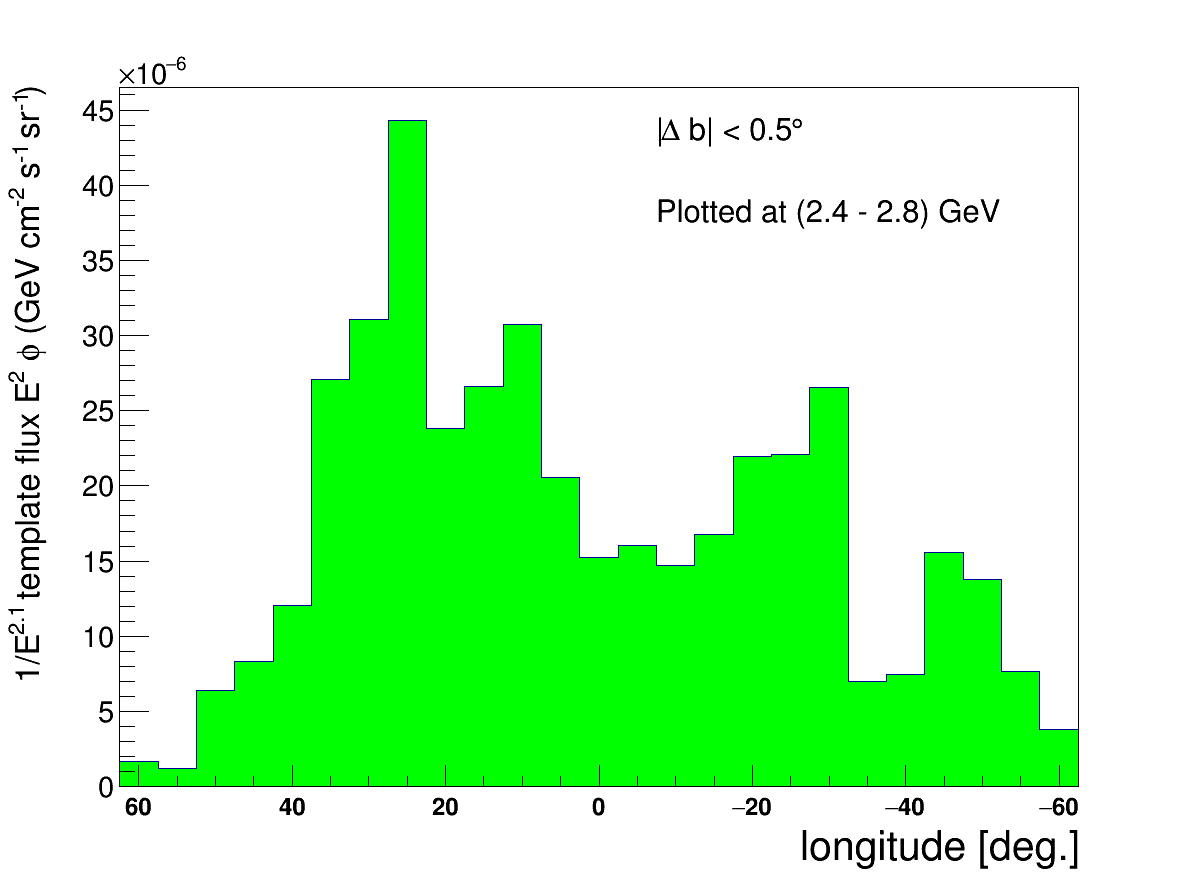}
\hspace*{0.01\textwidth}(a)\hspace*{0.45\textwidth} (b)
\caption[]{(a) Flux from the 1/E$^{2.1}$ template in the $l,b$ plane.  The contour corresponds to the morphology found by the authors of the discovery paper \cite{Su:2013}.  The units correspond to the bubble template unit in Fig. \ref{f1}d. (b)  Longitude distribution for the  flux from the $1/E^{2.1}$ template in the GP($|b|<0.5^\circ$). \label{f3}}
\end{figure}
Towards the GC the isotropic contribution is small and since its normalization is fixed mainly by regions outside our region-of-interest and the normalization is the same for all subcones, its normalization is fixed in our fit. Varying the isotropic component by $\pm 10\%$  from its nominal FST value for the diffuse class of the P7REP\_SOURCE\_V15 data worsened the fit.

Leaving $n_3$ and $n_2$ to be both free parameters in the fit leads to unstable fits because of the strong correlation of the BR and IC contributions at low energies. Therefore we take the ratio $n_3/n_2$  of the two templates   from the Dragon program for each sky direction. Varying this ratio by up to 30\% in each direction did not significantly change the results, nor the lower limit of 10\%  for IC or BR of the Dragon flux. A lower limit was imposed to stay away from  unphysical fits without IC or BR contributions.   Finally, there are only 3 free parameters for each subcone, namely  $n_1$, $n_2$ and $n_4$.

The energy spectrum in the Fermi bubbles is significant harder \cite{Fermi-LAT:2014sfa} and is well described by  a $1/E^{2.1}$ proton injection spectrum without any break, as shown in Fig. \ref{f1}d. The depletion below 1 GeV  originates from the kinematics of $\pi^0$ production.  
An example of a template fit with foregrounds only is shown in Fig. \ref{f2}a for a single subcone. One observes that the data below 10 GeV are well described by the sum of the foregrounds. However, above 10 GeV the data show an excess. This is the observation of the $1/E^{2.1}$ template in the GP.  Including this template leads to the fit shown in Fig. \ref{f2}b, which describes well the high energy excess.
%
\begin{figure}
\centering
\includegraphics[width=0.6\textwidth,clip]{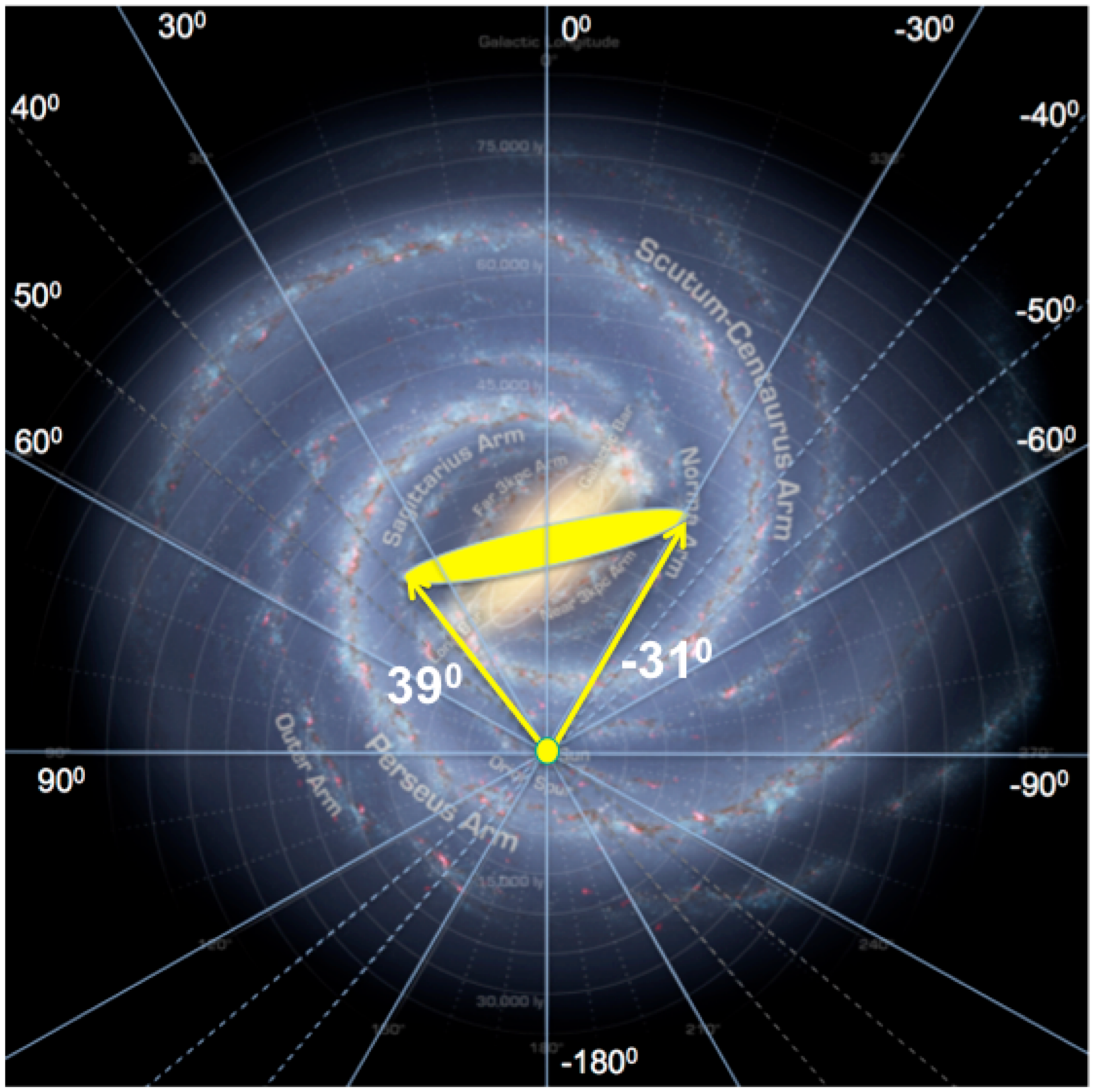}\vspace*{-1.5cm}
\caption[]{Sketch of the Milky Way with the central bar and spiral arms. The picture was adapted from  \cite{Churchwell:2009zz}). The "slim (yellow) ellipse" in the center indicates the angle and length of the region of the 1/E$^{2.1}$ template obtained from the endpoints of the bubble-like emission in Fig. \ref{f3}b.
 \label{f4}
}
\end{figure}

The  template fit discussed above was repeated for all  subcones. Surprisingly,  with only 3 free parameters for each direction ($n_1, n_2, n_4$) we can describe the whole gamma-ray sky with an excellent    $\chi^2$, except for some parts of the GP.  Here we needed to adjust the $\pi^0$ template towards regions with a high column density of MCs, as will be discussed in the last section.  
The correlation with the MCs  is demonstrated in Fig.  \ref{f2}c: the distribution in latitude of the  $1/E^{2.1}$ template is clearly narrower than the one of the $\pi^0$ template and corresponds to the latitude distribution from the column density of  MCs from Ref. \cite{Dame:2000sp}, shown by the inner solid line. The diffuse component has a  wider distribution, as shown by the outer dotted line, which was taken from the gas distribution, as implemented in the Dragon code. To get agreement between the width of the $\pi^0$ latitude distribution and the gas distribution it was important to take the $1/E^{2.1}$ template into account and to include the breaks in the injection spectra (see Sect. \ref{gce}), since otherwise no good fit is obtained and the $\pi^0$ latitude distribution is far too wide.

The flux from the $1/E^{2.1}$ template is  a factor $\approx$ 40 stronger in the GP than in the halo, as demonstrated by the narrow peak in Fig.  \ref{f2}d. In the GP the
intensity ratio R of the SRCs  and  $\pi^0$ production  from diffuse Galactic CRs was predicted \cite{2004ApJ...611...12B}:
 \begin{equation} R(E)=0.07\left(\frac{N_g^{SCR}}{N_g^{GCR}}\right)\left(\frac{T_p}{10^5 \ yr}\right)\left(\frac{E}{1\ GeV}\right)^{0.6}, \label{e3}\end{equation} where the first bracket takes into account the difference in gas densities in SNRs and the GP, the second bracket the limited confinement time $T_p$ of SCRs in SNRs and the third bracket the difference in energy dependence between the bubble template and the $\pi^0$ template.
From Fig.  \ref{f2}b one observes that this ratio becomes one for energies around 20 GeV, which requires that the product of the first two brackets is of the order of a few, which is the right order of magnitude, given that the gas density ratio itself is already of this order of magnitude.

\section{Morphology of the Fermi Bubbles}\label{morph}
The fitted values of  $n_4$, i.e. the flux from the $1/E^{2.1}$ contribution, in all sky directions are shown in Fig. \ref{f3}a, which obviously show a similar morphology as found  by  the authors of the discovery paper \cite{Su:2010qj}  (indicated by the surrounding line), but we find a much richer structure inside because of the  better spatial resolution with our method.  However, the surprise is a  strong bubble-like  emission in the  GP,  shown by the intense (red)  bar in the center of  Fig. \ref{f3}a. Its distribution as function of longitude is shown in Fig. \ref{f3}b, which reveals  a strong increase at $l=+39^\circ$  followed by a sharp decrease at $l=-31^\circ$. These angles are close  to the endpoints of the bar, as pictured in Fig. \ref{f4}. Assuming these to be the endpoints  and taking a distance between the GC and the Sun of $8.3\pm 0.4$ kpc \cite{Genzel:2010zy} the morphology of the bar is completely determined, as is apparent from the geometry shown in Fig. \ref{f4}. Using a conservative error for the directions to the endpoints of $\pm 1^\circ$ we find from a fit that the major bar axis makes an angle of  $77.7^\circ \pm 2.1^\circ$  with respect to the GC-Sun line and has a half length  of 5.9$\pm$0.1 kpc. This morphology is shown by the slim (yellow) ellipse in Fig. \ref{f4}. It has a  different angle than the bar indicated in the background, which has  $44^\circ\pm 10^\circ$ \cite{Churchwell:2009zz}, but this angle was found from a subset of stars in the first quadrant. We cannot claim that the complete $1/E^{2.1}$ contribution originates from the yellow bar region, since spiral arms before the bar may contribute as well.
Evidence for contributions from spiral arms comes from the fact that we see the bubble-like emission also at longitudes around  -50$^\circ$ as a separate peak in Fig. \ref{f3}b. This is exactly the region of the tangent point of the Scutum-Centaurus spiral arm, as shown in Fig. \ref{f4}.  
 
We interpret the combined findings as follows:  SNRs accelerate CRs  with a spectral index of 2.1 (on average) by diffusive shock wave acceleration, which in turn produce gamma-rays with such a hard spectrum as long as they are connected to the sources, the SCRs.   Towards the GC the global thermal and CR pressure can be high enough to blow a small fraction of the gas into the halo, either from the GP \cite{Everett:2007dw,Breitschwerdt:2008na} or from the inner GC \cite{Crocker:2011en,Crocker:2010qn}. This hot gas in the halo is inferred from the ROSAT X-ray data  \cite{Snowden:1997ze}, which was interpreted as advection of gas from the GP  \cite{BlandHawthorn:2002ij}.
\begin{figure}[t]
\centering
\includegraphics[width=0.84\textwidth,clip]{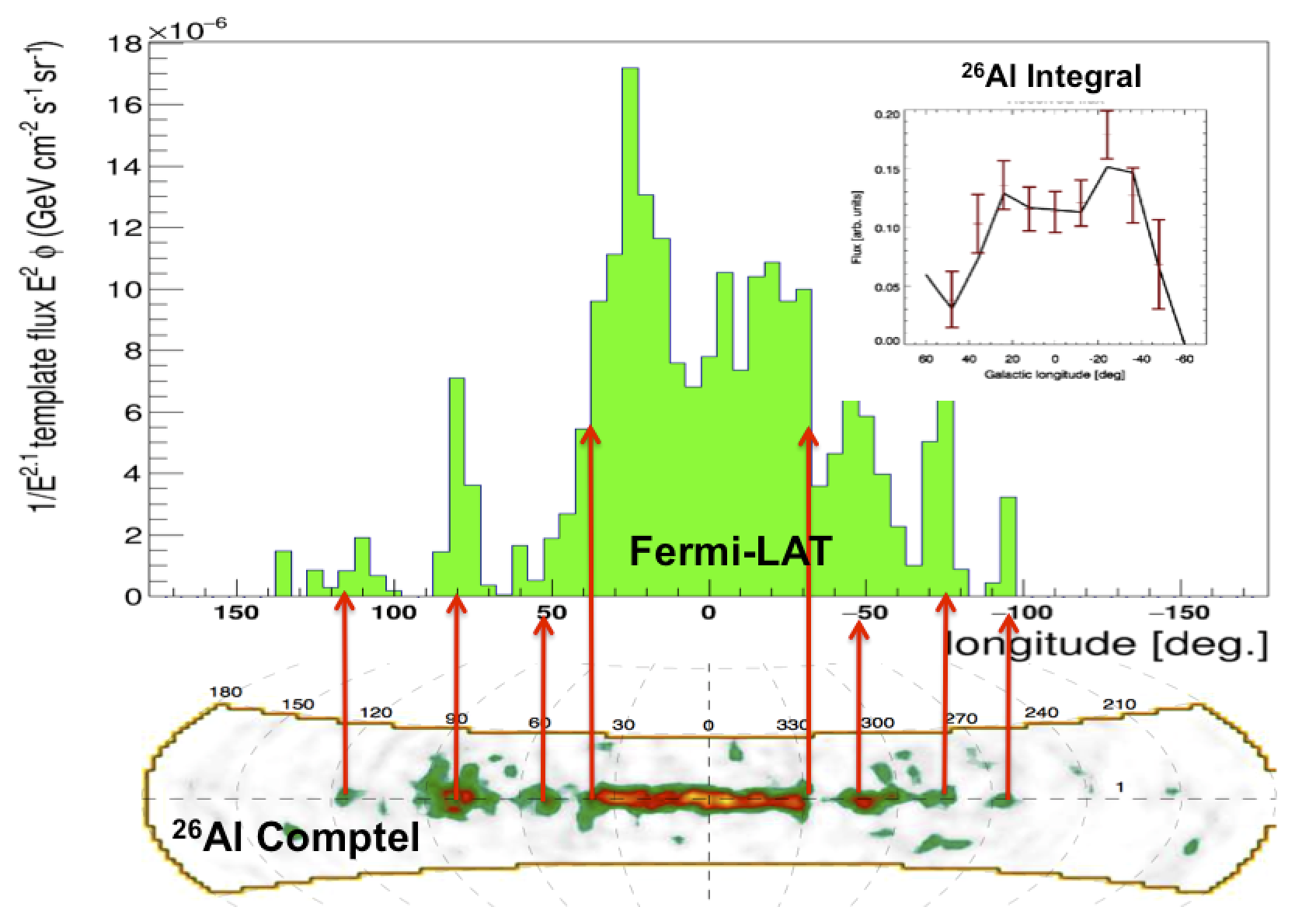} 
\caption[]{Morphology   of the 1/E$^{2.1} $ contribution (histogram) in the GP (|b|$<2^\circ$ in comparison with the morphology of the $^{26}$Al line (bottom graph  from the Comptel space mission \cite{Plueschke:2001dc} and the insert from the Integral space mission \cite{Kretschmer:2013naa}).
\label{f5}
}
\end{figure}
In such an advective environment the CRs can be trapped inside the plasma without an energy-dependent escape, so the  SCRs will still correspond to the 1/$E^{2.1}$  spectrum at high latitudes
or the CRs are accelerated in the shocked gas from the outflow \cite{Crocker:2014fla}. The spectral index of the locally observed proton spectrum is around 2.8 and the softening from 2.1 to 2.8 is attributed to diffusion, simply because high energy protons escape faster from the sources and the Galaxy \cite{Strong:2007nh}. 
\section{Energetics}\label{ener}
 The luminosity of the 1/E$^{2.1}$ emission in the halo  between 1 and 100 GeV for $10^\circ<|b|<55^\circ$ and  $|l|<30^\circ$ corresponds to $5.6\pm0.3(fit)\pm0.9(sys)\cdot10^{37}$   erg/s. Here we followed  the calculation  in the discovery paper by \cite{Su:2010qj}, who found  $4\cdot10^{37}$ erg/s  for the halo bubbles without giving an error. The first error originates from the fit, while the second error originates from the error in the spectral index, which is the dominant systematic uncertainty because the normalization of the bubble template is most sensitive to the high energy tail in the data, so the extrapolation to lower energies depends on the spectral index. The bubble luminosity increases to  $8.7\pm0.4(fit)\pm1.5(sys)\cdot10^{37}$ erg/s, if we decrease the  latitude limit to  $1.5^\circ$, while the 1/E$^{2.1}$ emission in the GP ($|b|<1.5^\circ$) has a luminosity of $6.9\pm0.6(fit)\pm1.1(sys)\cdot10^{37}$ erg/s. This can be compared with the hadronic energy release from SNRs in the inner Galaxy.  We expect for the luminosity in gamma-rays  between 1 and 100 GeV:  $E_\gamma=\epsilon_{CR}\  \epsilon_\gamma\  \epsilon_{(1-100)} \ \epsilon_{SCR/(SCR+GCR)}\ E_0$ erg/s. Here  $\epsilon_{CR}$ is the fraction of the SNR mechanical power $E_0$ into hadronic CR energy,  $ \epsilon_\gamma$ is the energy converted into gamma-rays, $ \epsilon_{(1-100)}$ is the fraction of gamma-ray energy between 1 and 100 GeV, and $\epsilon_{SCR/(SCR+GCR)}$ is the fraction of energy into SCRs.
Hillas  \cite{Hillas:2005cs} estimates a SNR rate of 1 SNR/century in the inner region ($R<4$ kpc) corresponding to $E_0=10^{51}\ erg/3.10^9\ s=3.3\cdot 10^{41}$ erg/s. Using   $\epsilon_{CR}\ \epsilon_\gamma\  \epsilon_{1-100}\ \epsilon_{SCR/(SCR+GCR)}$=0.16 x 0.1 x 0.3 x 0.3=0.0009 we find  $E_\gamma=4.8\cdot10^{38}$ erg/s  for the gamma-ray energy from SCRs to be compared with a summed bubble-like emission in the halo and disc of $E_\gamma^{SCR}=1.6\cdot10^{38}$ erg/s. This is reasonable agreement given the large uncertainties involved.
We obtained $\epsilon_{SCR/(SCR+GCR)}=0.3$ from the energies in the $n_4$ and $n_1$ components in the fit, while the other efficiencies were estimated as in Ref. \cite{Hillas:2005cs}.
\begin{figure}[]
\centering\hspace*{2mm}
\includegraphics[width=0.44\textwidth,height=0.33\textwidth,clip]{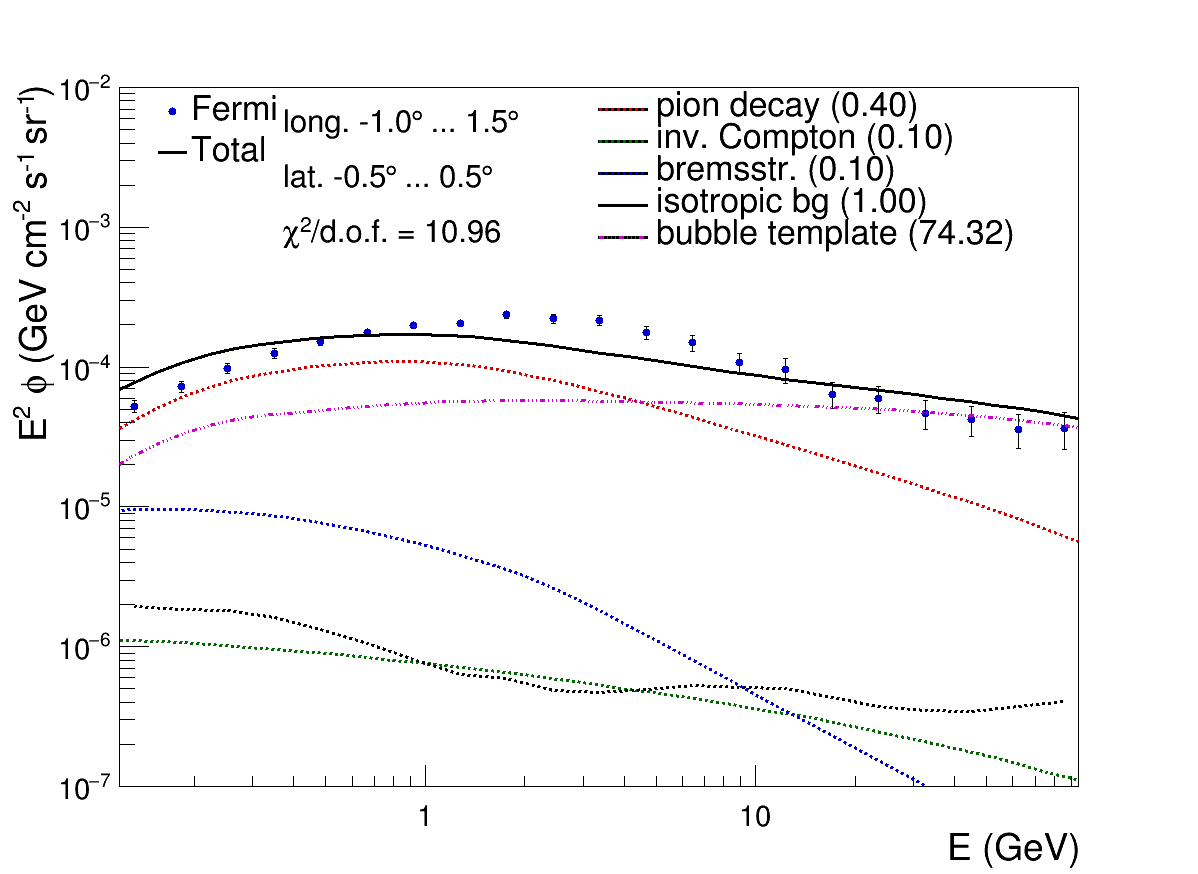}
\hspace*{5mm}
\includegraphics[width=0.44\textwidth,height=0.33\textwidth,clip]{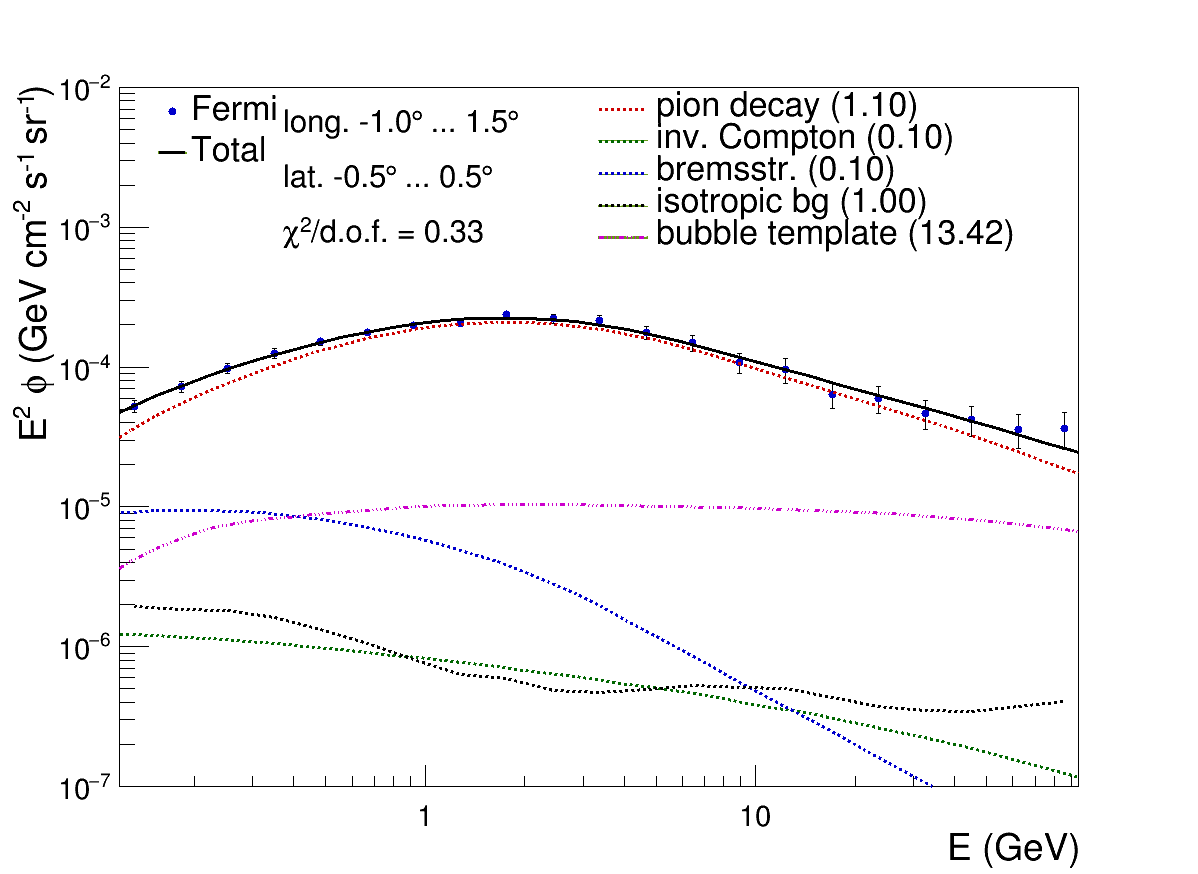}
\hspace*{0.03\textwidth}(a)\hspace*{0.45\textwidth} (b)\\
\hspace*{3mm}
\includegraphics[width=0.44\textwidth,height=0.33\textwidth,clip]{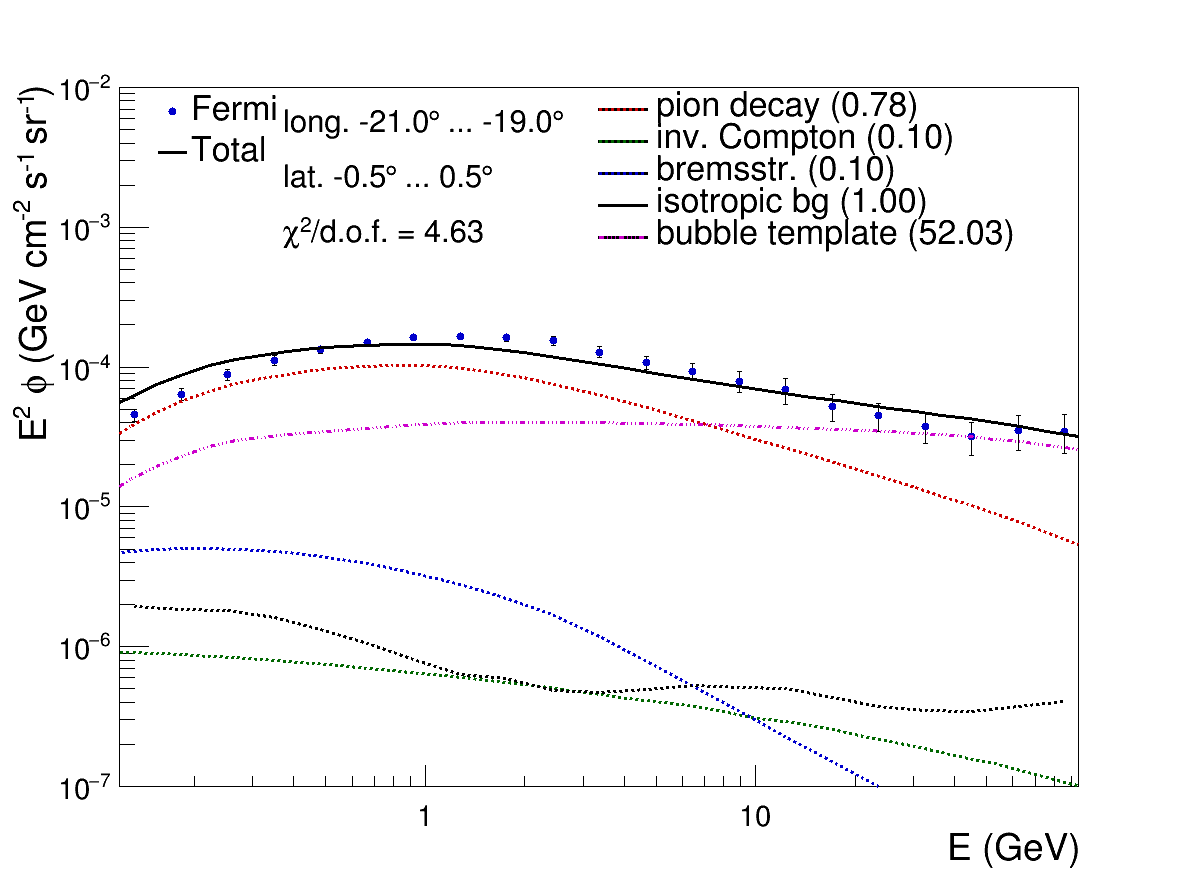}
\hspace*{5mm}
\includegraphics[width=0.44\textwidth,height=0.33\textwidth,clip]{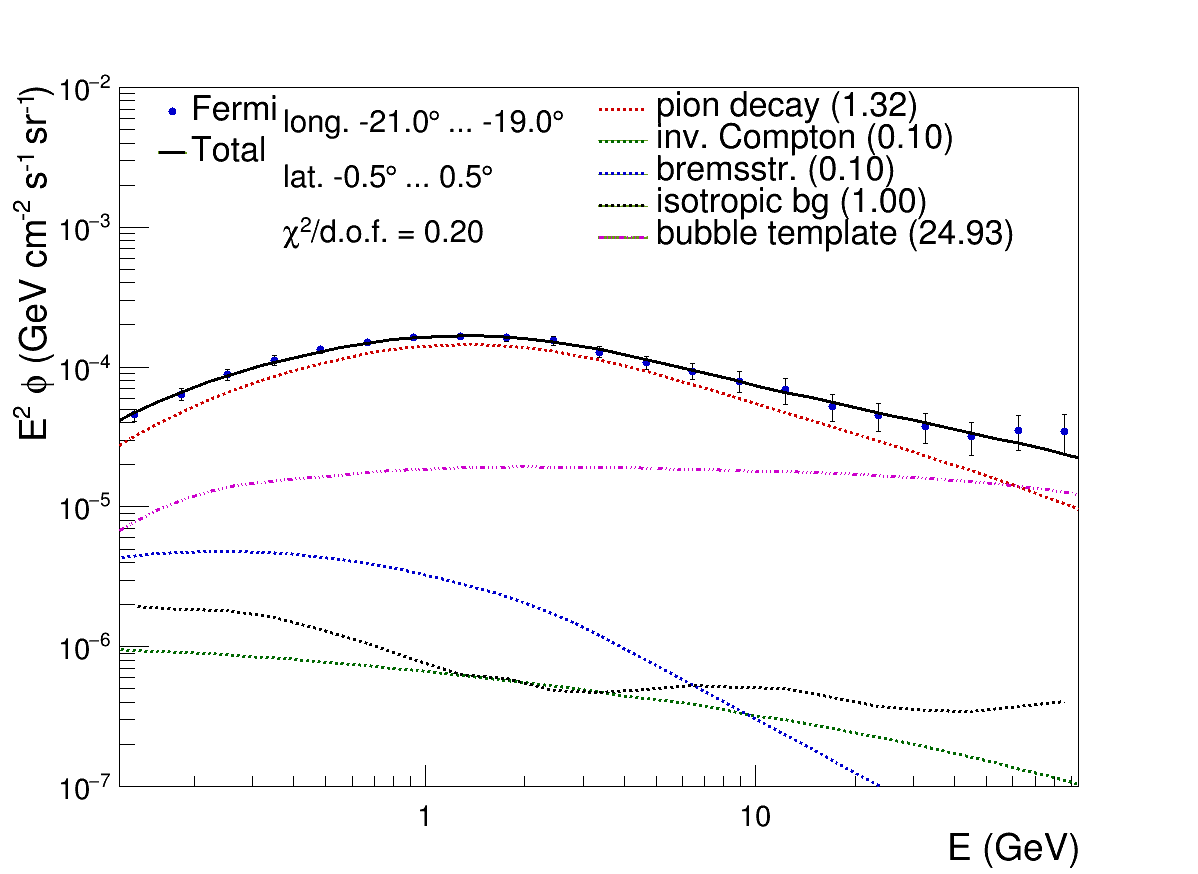}
\hspace*{0.03\textwidth}(c)\hspace*{0.45\textwidth} (d)  
\hspace*{3mm}
\includegraphics[width=0.44\textwidth,height=0.33\textwidth,clip]{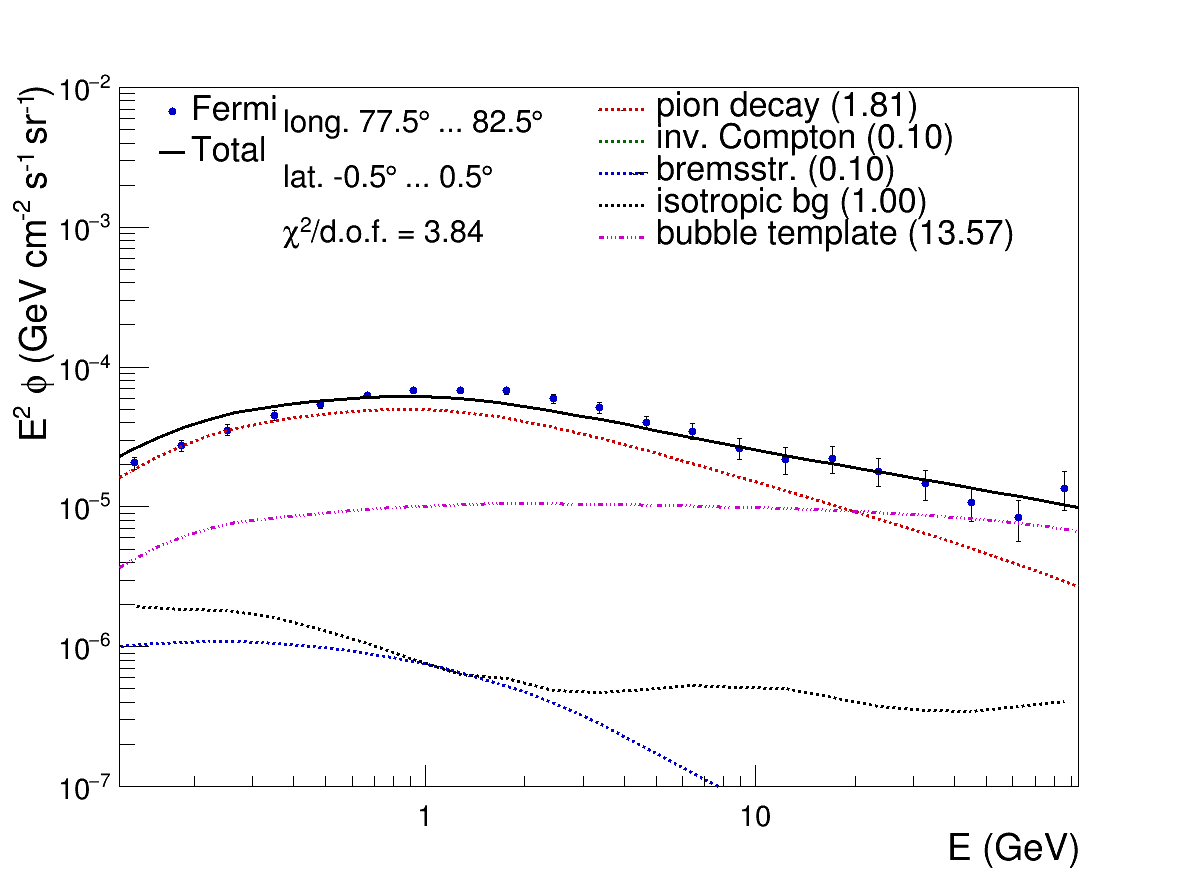}
\hspace*{5mm}
\includegraphics[width=0.44\textwidth,height=0.33\textwidth,clip]{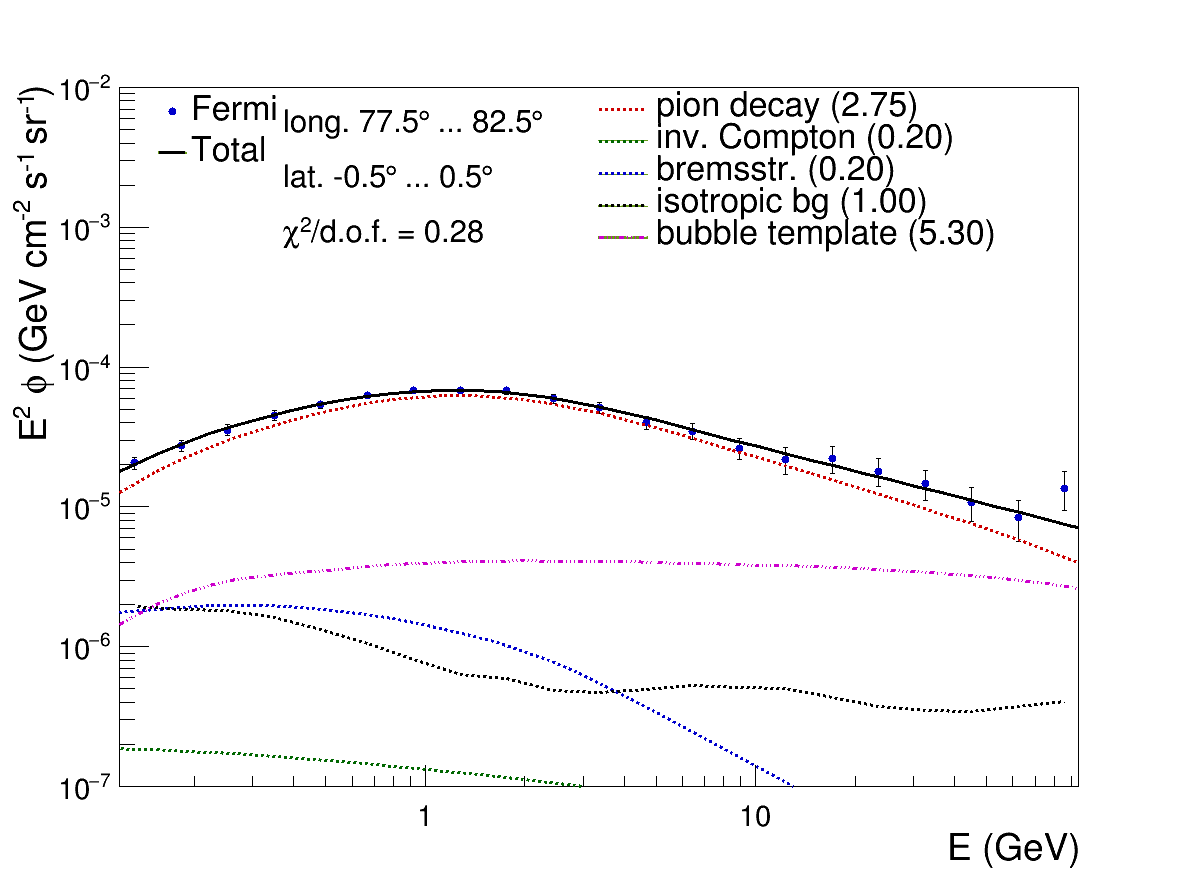}
\hspace*{0.03\textwidth}(e)\hspace*{0.45\textwidth} (f)  
\caption[]{Left:  some examples of the template fit in regions with strong $^{26}$Al production  without using a break  in the injection spectrum.  The few GeV excess in the data is clearly visible at  longitudes even far away from the GC ($l=80^\circ$).  A break is needed in all directions with strong $^{26}$Al production, see text.   Right: as on the left, but the excess  disappears for a break in the injection spectra, as indicated in Fig. \ref{f7}. The higher the MCs density, the higher the break. Note that the break in the Pamela data is thought to originate from stellar winds and energy losses, so in regions of dense MCs it is not unreasonable to have deficit of low energy protons.  The numbers in brackets in the legends indicate the normalization of the described contribution with respect to the Dragon prediction or the strength of the bubble template in units of the solid curve in Fig. \ref{f1}d.
\label{f6}
}
\end{figure}
\section{Comparison with $^{26}$Al sky maps}\label{26al}
The radioactive isotope  $^{26}$Al is synthesized in stars and expelled into the interstellar media by winds or supernovae explosions. It can be observed by its characteristic 1.809 MeV gamma-ray line emitted in the decay (lifetime 7.10$^5$ years). Given the small $\beta=v/c$ factor in combination with its short lifetime the element is expected to decay close to the source and hence to be a good tracer of SNRs. Although other stars do contribute, SNRs are expected to be the dominant source. Since the SCRs trace SNRs as well, we should expect a close correlation between the skymaps of the 1.809 MeV signal of $^{26}$Al and the multi-GeV hard component in the Fermi gamma-ray data. This is indeed the case,  as can be seen from a comparison of the longitude distributions of the $1/E^{2.1}$ and $^{26}$Al fluxes  \cite{Plueschke:2001dc,Kretschmer:2013naa} in Fig. \ref{f5}. 
Note that the spatial resolution of the   Fermi LAT instrument is around 0.5$^\circ$, while for the $^{26}$Al data from Comptel it is about 1.6$^\circ$.  
%

\section{The GC GeV Excess (GCE)}\label{gce}
\begin{figure}[t]
\centering\hspace*{2mm}
\includegraphics[width=0.5\textwidth,height=0.37\textwidth,clip]{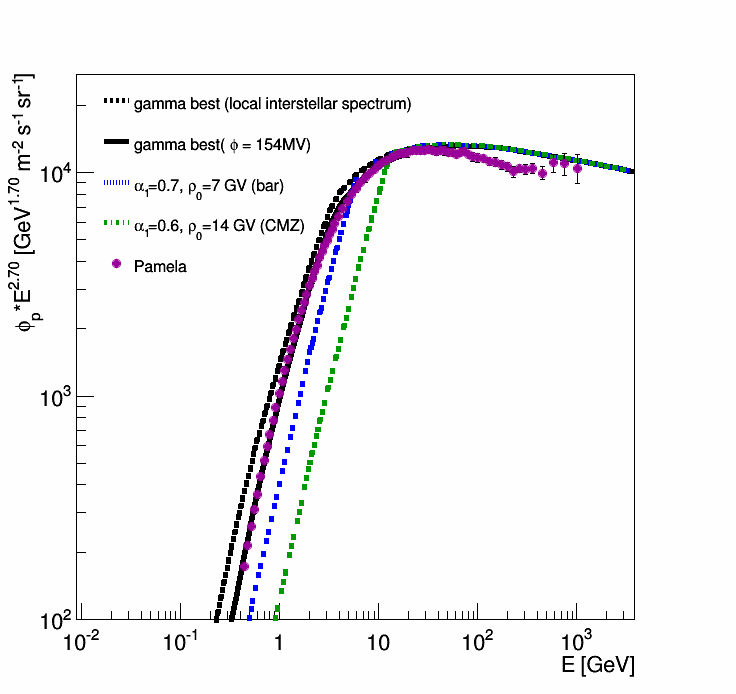}\hspace*{1mm}
\includegraphics[width=0.47\textwidth,height=0.33\textwidth,clip]{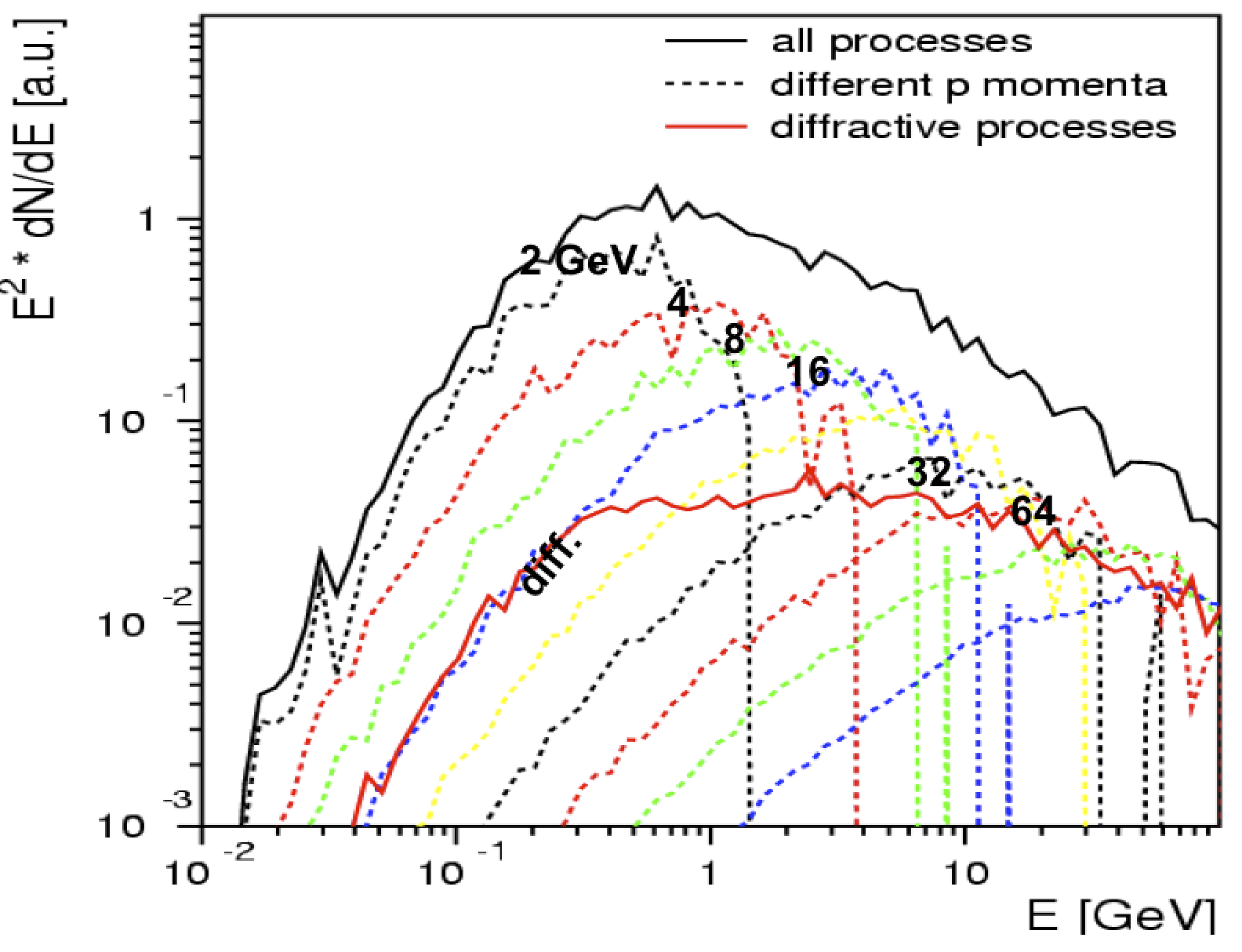}
\hspace*{0.03\textwidth}(a)\hspace*{0.4\textwidth} (b)
\caption[]{(a): Several proton spectra describing best the gamma-ray data in various regions: black dashed line for most of the sky, i.e.  outside the regions of MCs. The solid line corresponds to the dashed line after applying a solar modulation of 154 MeV, which can be compared  with the Pamela data (purple circles). The dashed lines with depletion at low rigidities are the spectra with a  7 (blue) and 14 GV (green) break, respectively. Breaks between 7 and 14 GV are needed to describe the gamma-ray spectra in MCs (see text).  (b): the gamma-ray spectrum for various slices of the proton spectrum. Note that the maximum photon energy is always below the proton energy. One observes clearly a shift of the maximum flux, if the low energy protons are suppressed by a break as indicated in (a).
\label{f7}
}
\end{figure}

Towards the GC the template fit gets considerably worse, since the data show an apparent excess at a few GeV, as demonstrated in Fig. \ref{f6}a. This excess turns out to be present everywhere in the GP, where  the  $^{26}$Al production is strong. A few examples are shown in Figs. \ref{f6}c and \ref{f6}e.  This excess can be removed by a break in the injection spectra, leading to a shape of the proton spectra in Fig. \ref{f7}a.  The curve at the left side shows the interstellar spectrum best describing the gamma-ray sky outside the GC and bubble regions, i.e. for $|l|>40^\circ$. This tuning is called "gamma-best".  After solar winds are taken into account the "gamma-best" proton spectrum resembles  the observed local  proton spectra, as shown by the Pamala data points. Energy losses and stellar winds deplete the observed spectra at low energies. It is not unreasonable to expect that these effects deplete the lower energy part of the proton spectra even more in the environment of dense MCs. Such a deficit can be parametrized by a break in the injection spectrum, as shown by the dashed lines in Fig. \ref{f7}a for a break at 7 and 14 GV, respectively.  Such breaks immediately remove the apparent gamma-ray excess, as shown on the right hand side of Fig. \ref{f6}. 
This can be understood from Fig.  \ref{f7}b, which shows how a deficit  of low energy protons in the parent proton population shifts the maximum in the gamma-ray spectrum from below 1 GeV to a few GeV.
A  break at 7 GV is needed for a good fit towards the tangent point of the spiral arm at  a longitude of -50$^\circ$, but in directions with a higher column density of MCs (defined by a higher flux of $^{26}$Al production)  the needed break shifts to slightly higher rigidities: 8 GV for the "Cygnus" region at $77.5^\circ<l<82.5^\circ$, 9 GV for  the direction of the bar ($-31^\circ<l<39^\circ$) and 14 GV towards the Central Molecular Zone (CMZ) at $-1^\circ<l<1.5^\circ$ \cite{Tsuboi:1999,Jones:2011wb}.  The density in the CMZ is several orders of magnitude higher than the averaged gas density. 
\begin{figure}[t]
\centering\hspace*{2mm}
\includegraphics[width=0.44\textwidth,height=0.33\textwidth,clip]{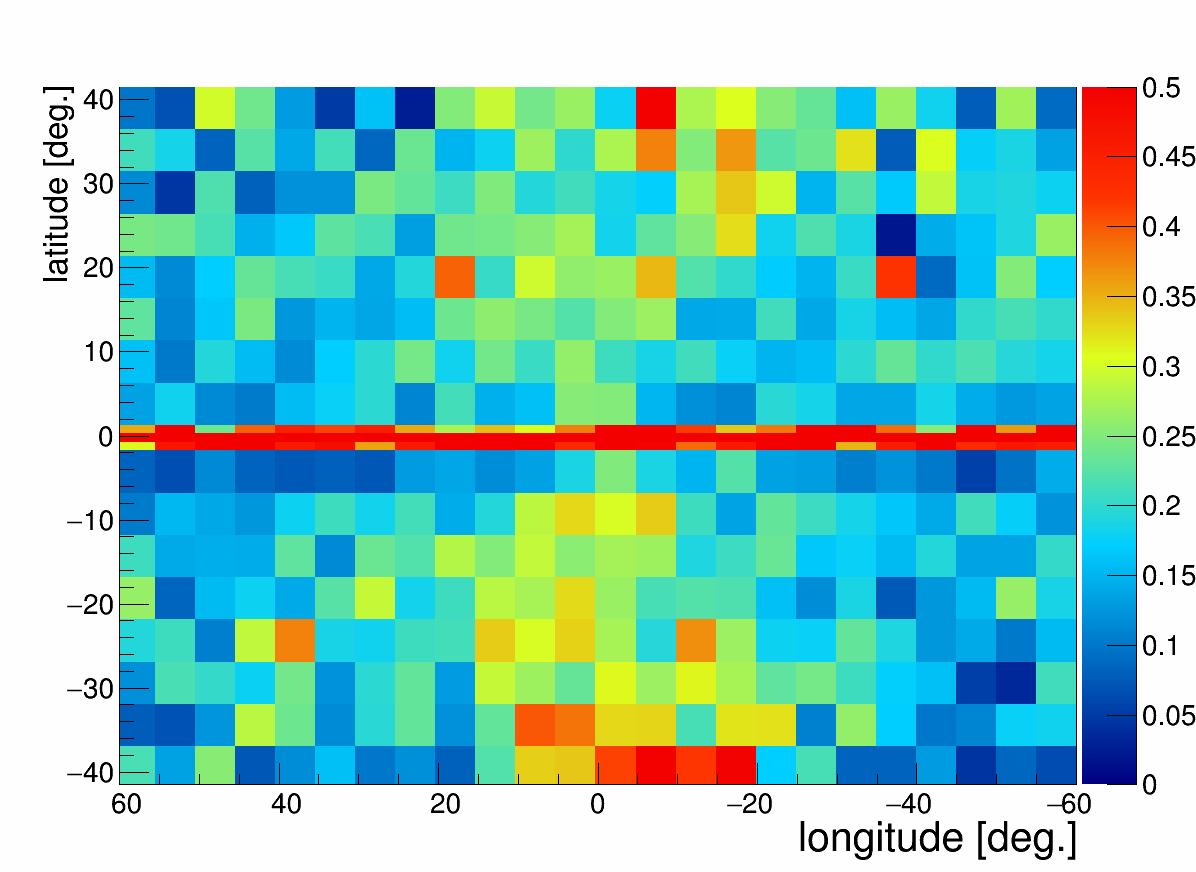}
\hspace*{5mm}
\includegraphics[width=0.44\textwidth,height=0.33\textwidth,clip]{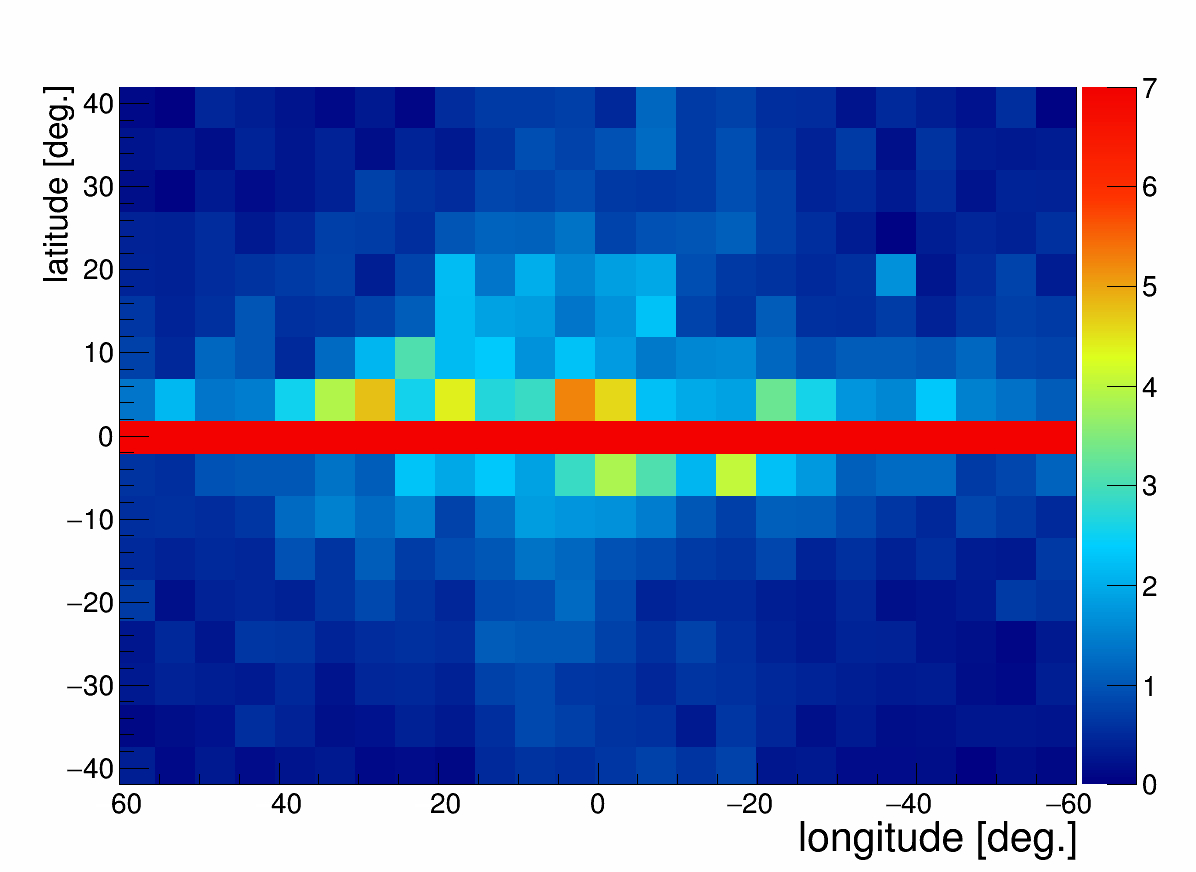}
\hspace*{0.01\textwidth}(a)\hspace*{0.5\textwidth} (b)\\ \hspace*{3mm}
\includegraphics[width=0.44\textwidth,height=0.33\textwidth,clip]{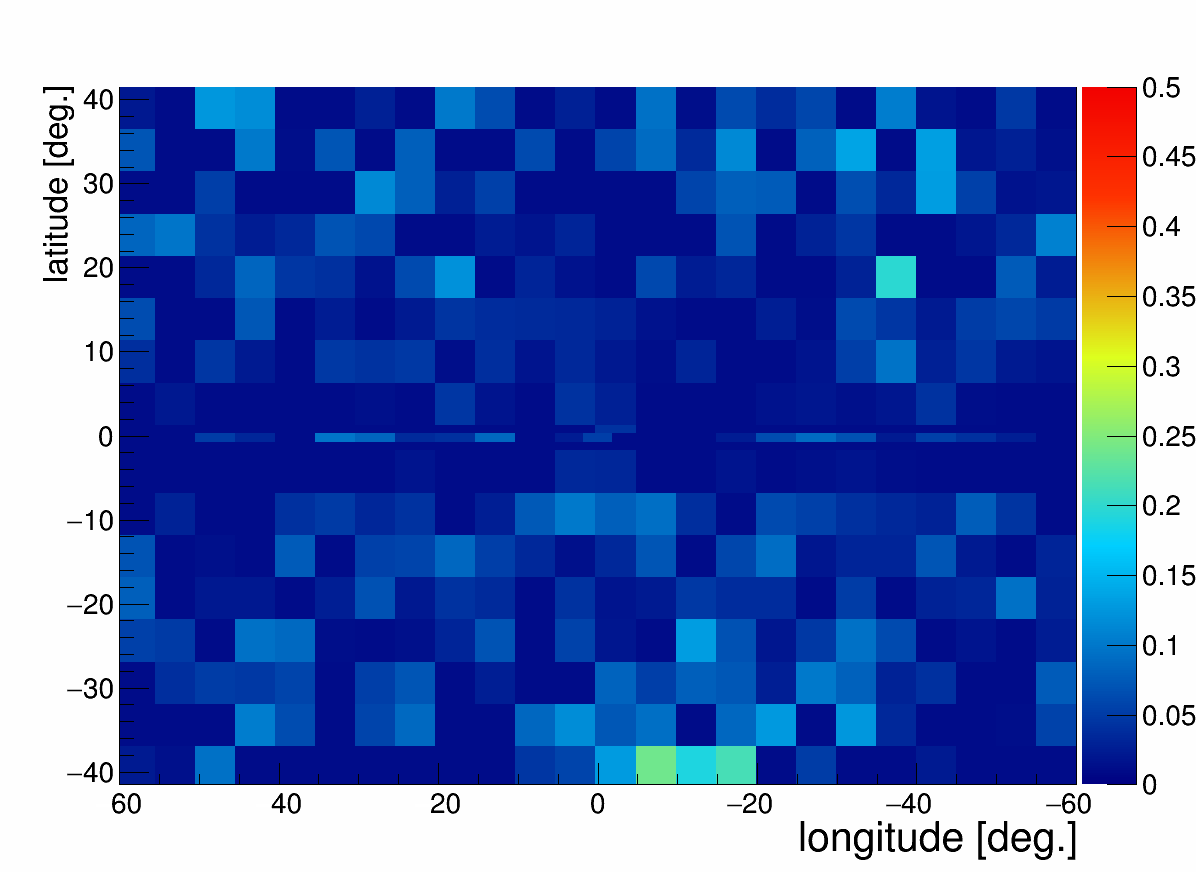}
\hspace*{5mm}
\includegraphics[width=0.44\textwidth,height=0.33\textwidth,clip]{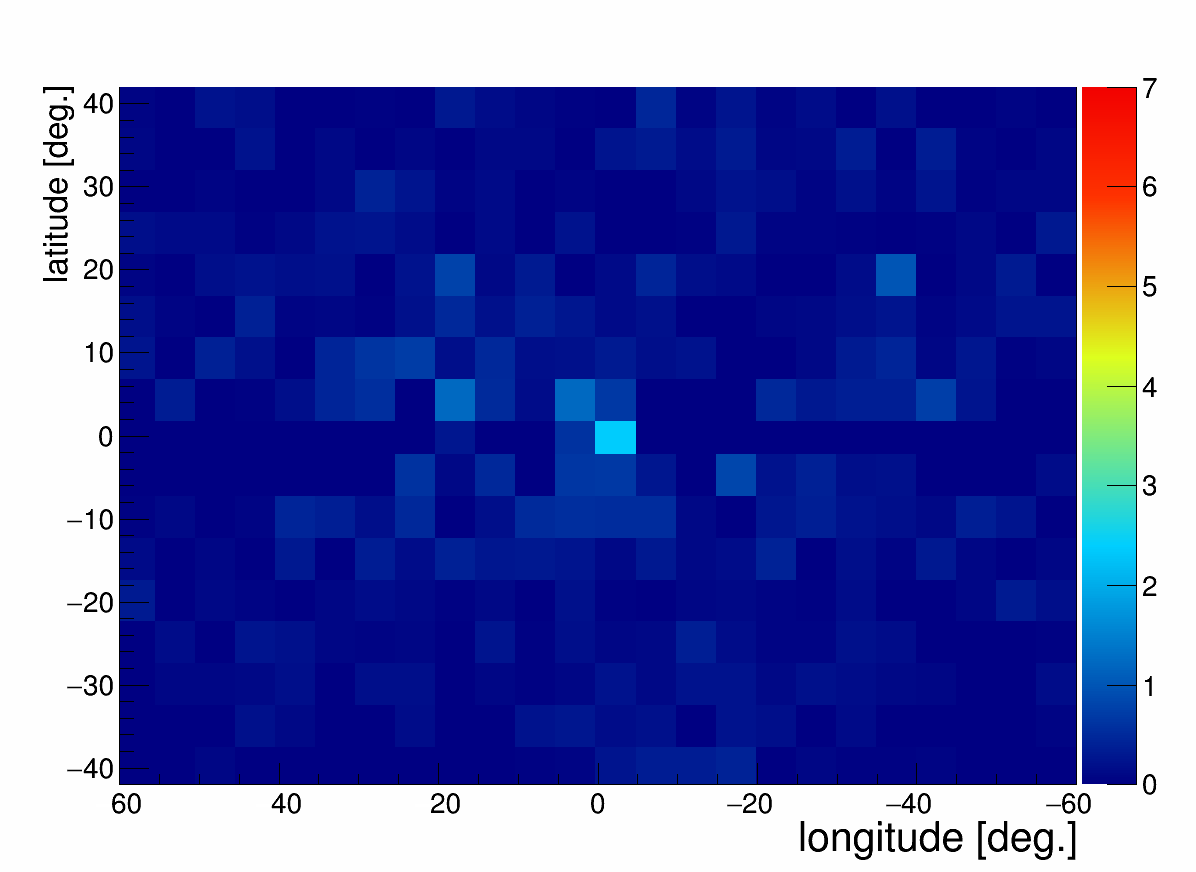}
\hspace*{0.01\textwidth}(c)\hspace*{0.5\textwidth} (d)
\caption[]{(a) Differences  between the fit and the data (for energies around 2.4), if the templates are based on the injection spectra from the locally observed CRs. The colour coding indicates the relative difference between data and the best fit value for values between 0 and 50\%. One observes a clear excess in the bubble region of about 30\% and the excess near the GP varies between 60 and 120\%. (b) As in (a), but instead of the relative difference the absolute difference is plotted in units of $10^{-6}$ $\rm GeV ~cm^{-1}~ s^{-1} sr^{-1}$. Because of the high flux in the GP the large relative error causes a large absolute excess (varying from 20 at |l|=60$^\circ$ to 140 at the GC in units of $10^{-6}$ $\rm GeV ~cm^{-1}~ s^{-1} sr^{-1}$)  and the excess at higher latitudes is just an artefact of the l.o.s crossing the GP with this large excess combined with a higher absolute flux towards the GC. (c) As in (a) but after optimizing the injection spectra of protons and including the $1/E^{2.1}$ template, which removes the excess in all regions, especially  the strong excess in the GP, which is largely responsible for the excess seen in (b). (d) The absolute excess disappears after the excess in the GP  disappears with the tuning in (c).
\label{f8}
}
\end{figure}

How does our analysis compare with other analyses discussing the GCE?
All previous analyses mask the region in the GP, since the diffuse gamma-ray emissivity is here poorly known. However, {\it it is a misconception}, if one believes not to be sensitive to the GP in this case, since the integral along the l.o.s. of the gamma-rays always crosses the GP and any excess in the GP will show up as an excess in whatever direction. Since the column density in the GP increases with decreasing latitude, the effect becomes strongest at low latitudes.  This is demonstrated in Fig.  \ref{f8}a, which shows the relative residuals, defined as (data-model)/data. Here the model is the predicted flux for templates from nuclei injection indices without a break. One observes about 30\% deviations in the regions of the Fermi Bubbles (even at the low energy of 2.4 GeV, for which the residuals are shown), but the deviations in the GP are much larger and vary between 60 and 120\%.  If we plot these deviations as an absolute excess, we observe the famous excess in the l,b-plane towards the  GC (see Fig. \ref{f8}b). It is largest in the GP because of the large flux there combined with the large relative deviation (60-120\%), but it is  visible up to latitudes of 20$^\circ$. The latitude dependence is easily explained by the increasing column density of the GP to the integral along the l.o.s., as discussed above. But why the excess is strongest at small longitudes? The answer is simple: the flux towards the GC is increasing (because of the higher gas density) and the excess is a certain percentage, so the absolute value of the excess increases with the absolute flux, i.e. with decreasing longitude.
If we remove the excesses in the GP by retuning  the injection spectrum (via the breaks in the  regions of strong , as discussed above) and add the $1/E^{2.1}$ template, the observed and calculated fluxes are in excellent agreement in the whole sky  (see Fig. \ref{f8}c) and the "pseudo-GC" excess disappears, as shown in Fig. \ref{f8}d. Note that a good fit requires to take the $1/E^{2.1}$ template into account at all latitudes, even at zero degree, as is apparent from Fig. \ref{f3}a.
We only optimized the dominant hadronic emission in the GP, which was sufficient to eliminate  the GCE.  This clearly proves that the apparent GCE is due to the uncertain emissivity in the GP, not in the GC.

\section{Summary}\label{conc}
In summary, with our template fitting method we confirm the bubbles in the halo, but find in addition  strong bubble-like emission in the GP.  Both, in the halo and the GP, the spectral index of the hard gamma-ray component corresponds to a proton injection index of 2.1 without a break , as expected for "freshly" accelerated CRs inside the sources (SCRs). The narrow latitude distribution of the 1/E$^{2.1}$ contribution shows that it originates mainly from the region of the molecular clouds, as expected for SCRs. In addition, the longitude distribution of the 1/E$^{2.1}$ contribution coincides with the emission of $^{26}$Al, which is presumed to be a tracer of SNRs \cite{Prantzos1996}. The sharp edge of the bubble-like emission in the GP at longitude l=-31$^{\circ}$ in Fig. \ref{f3}b) coincides with the endpoint of the Galactic Bar, while the bump at  l=-50$^{\circ}$ coincides with the tangent point of the nearby Scutum-Centaurus  arm (Fig. \ref{f4}).  These three observations (agreement with predicted SCR spectrum, correlation with SNRs via the correlation with the $^{26}$Al line  and correlation with a nearby spiral arm are  strong arguments for associating the 1/E$^{2.1}$ contribution with SCRs.

Why the SCRs and the Fermi Bubbles have the same 1/E$^{2.1}$ spectrum?
This is most easily explained, if one assumes the bubbles to be outflows of the star-forming region near the GC, where the combined thermal- and CR pressure is high enough to overcome the gravitational pull. If the protons are trapped inside the plasma,  their spectra are not softened by energy-dependent escape. This  leads to the same hard  1/E$^{2.1}$ gamma-ray spectra inside the sources and in the bubbles.
Alternatively, the protons may be accelerated in the shock wave of the outflow, thus having the same spectrum as the SCRs \cite{Crocker:2014fla}, which are accelerated in shock waves as well.

 Finally, we showed that the famous few GeV excess towards the GCE also occurs in the GP (GPE) and is strongly correlated with the $^{26}$Al flux, just as the  1/E$^{2.1}$ contribution. As discussed above, the latter contribution was interpreted as the contribution from "freshly" accelerated CRs inside sources (SCRs). In contrast, the few GeV excess was interpreted as the contribution from "old" CRs obtaining a deficit at low rigidity in the dense environment of MCs, characterized by higher energy losses and stronger stellar winds. Since the SCRs and dense MCs are both traced by  $^{26}$Al  production, it comes as no surprise that the few GeV excess (from "old" CRs) and the   1/E$^{2.1}$ contribution (from "fresh" SCRs) have an identical morphology, namely the one from the $^{26}$Al flux. The reason why the GPE shows up as a GCE  is simply that the l.o.s towards the GC cross the GP.  If we retune the injection spectra in the GP to correctly describe the  gamma-rays for regions with a high column density of  MCs, the GCE indeed disappears, as expected if the GCE is mainly an artefact of the GPE.

\acknowledgments
Support from the Deutsche Forschungsgemeinschaft  (DFG, Grant BO 1604/3-1)  is warmly  acknowledged. We thank Roland Crocker, Francesca Calore, Daniele Gaggero  and Christoph Weniger for helpful discussions. We are grateful to the Fermi scientists, engineers and technicians for collecting the Fermi data and the Fermi Science Support Center for providing the software and strong support for  guest investigators.

\providecommand{\href}[2]{#2}\begingroup\raggedright\endgroup

\end{document}